\documentclass[%
 aip,
 jmp,%
 amsmath,amssymb,
 preprintnumbers,
 reprint,%
 twocolumn,
nofootinbib
]{revtex4}

\usepackage{graphicx}
\usepackage{dcolumn}
\usepackage{bm}

\usepackage{epstopdf}

\graphicspath{{./fig/}{./png/}}
\newcommand{\Tab}[1]{Table~\ref{#1}}
\newcommand{\Fig}[1]{Fig.~\ref{#1}}
\newcommand{\bra}[1]{\langle #1\rangle}

\newcommand{\EQ}{\begin{equation}}
\newcommand{\EN}{\end{equation}}
\newcommand{\dd}{{\rm d} {}}

\newcommand{\uu}{\bm{u}}
\def\Rey{\mbox{\rm Re}}
\def\Co{\mbox{\rm Co}}
\def\nuT{\nu_{\rm T}}
\def\urms{u_{\rm rms}}
\def\kf{k_{\rm f}}

\begin{document}

\preprint{Phys.\ Rev.\ E, Vol.\ 93, 033125 (2016); NORDITA-2015-118}

\title[Flow generation by inhomogeneous helicity]{Large-scale flow generation by inhomogeneous helicity}

\author{N. Yokoi}
\altaffiliation[]{Guest researcher at the National Astronomical Observatory of Japan (NAOJ) and the Nordic Institute for Theoretical Physics (NORDITA)}
\email{nobyokoi@iis.u-tokyo.ac.jp}
\affiliation{Institute of Industrial Science, University of Tokyo, Tokyo, Japan}

\author{A. Brandenburg}%
\affiliation{
Nordita, KTH Royal Institute of Technology and Stockholm University, SE-10691 Stockholm, Sweden
}\affiliation{
Department of Astronomy, Stockholm University, SE-10691 Stockholm, Sweden
}\affiliation{
JILA and Department of Astrophysical and Planetary Sciences, 
Laboratory for Atmospheric and Space Physics, University of Colorado, Boulder, CO 80303, USA
}

\date{28 March 2016}             

\begin{abstract}

The effect of kinetic helicity (velocity--vorticity correlation) on turbulent momentum transport is investigated. The turbulent kinetic helicity (pseudoscalar) enters the Reynolds stress (mirrorsymmetric tensor) expression in the form of a helicity gradient as the coupling coefficient for the mean vorticity and/or the angular velocity (axial vector), which suggests the possibility of mean-flow generation in the presence of inhomogeneous helicity. This inhomogeneous helicity effect, which was previously confirmed at the level of a turbulence- or closure-model simulation, is examined with the aid of direct numerical simulations of rotating turbulence with non-uniform helicity sustained by an external forcing. The numerical simulations show that the spatial distribution of the Reynolds stress is in agreement with the helicity-related term coupled with the angular velocity, and that a large-scale flow is generated in the direction of angular velocity. Such a large-scale flow is not induced in the case of homogeneous turbulent helicity. This result confirms the validity of the inhomogeneous helicity effect in large-scale flow generation and suggests that a vortex dynamo is possible even in incompressible turbulence where there is no baroclinicity effect. 

\end{abstract}

\maketitle

\section{Introduction\label{sec:1}}
Vortical structures are ubiquitously observed in hydrodynamic and
magnetohydrodynamic phenomena. The genesis of cyclones (typhoon,
hurricane, tornado, etc.) is one of the open problems in
atmospheric science. Small-scale vortical structures in turbulence are
considered to be the cause of large-scale magnetic fields in geo- and
astro-physical objects \cite{par1955,bra2005}. Recently the importance of
the large-scale vortical motions in the dynamo process has been discussed
\cite{yok2013}. It was also pointed out that a swirling structure may
play an important role in channeling energy from the lower photosphere into
the upper solar atmosphere \cite{wed2012}.
To understand these processes better, the large-scale vorticity
generation mechanisms in turbulence should be studied from various
viewpoints.

Turbulent kinetic helicity resulting from velocity--vorticity fluctuation
correlation represents the topological or structural properties of turbulence.
It has been noted that in the presence of helicity, a suppression
of turbulent energy transfer may occur due to the topological constraint
related to the possible conservation of kinetic helicity \cite{mof1992}. In the context
of local turbulent transport, helicity is expected to play some role
in momentum-transport suppression \cite{yok1993}. This is in contrast
to the turbulent or eddy viscosity, which is expressed in terms of the
turbulent energy (intensity information of turbulence), and represents
an enhanced transport due to turbulence.

	In homogeneous isotropic turbulence studies, helicity has been discussed in the context of a relation to the inverse energy cascade from larger to smaller wavenumbers, or reduction of the turbulent energy cascade \cite{bri1973,kra1973}. Using a numerical simulation of a variant of the Eddy-Damped Quasi-Normalized Markovian (EDQNM) approximation closure equations, Andr\'{e} and Lesieur \cite{and1977} showed that helicity influences the energy transfer rate of turbulent energy towards small scales. Their results showed that helicity suppresses the energy transfer to the small scales in the early stage of evolution, but once the inertial range has been established, such suppression effects disappear. As for the recent studies on the inverse energy cascade and helicity in three-dimensional rotating and stratified turbulence, see a series of papers by Pouquet, et al.\ \cite{pou2013,mar2013a,mar2013b}.

	The relationship between the helicity density and the dissipation rate has been investigated in several homogeneous and pipe flow geometries. Using Direct Numerical Simulations (DNSs) of the Navier--Stokes equation in channel and Taylor--Green vortex flows, Pelz {\textit{et al.}}\ \cite{pel1985} examined the local helicity density $\langle {{\bf{u}} \cdot {\mbox{\boldmath$\omega$}}} \rangle$ [${\bf{u}}$: velocity, $\mbox{\boldmath$\omega$} (= \nabla \times {\bf{u}})$: vorticity]. They found that the alignment between the velocity and vorticity is stronger in the region where the dissipation rate is smaller. 

	However, detailed numerical results in several homogeneous flows and fully developed turbulent channel flow by Rogers and Moin \cite{rog1987} showed no correlation between the relative helicity density $\langle {{\bf{u}} \cdot \mbox{\boldmath$\omega$}} \rangle / (|{\bf{u}}| |\mbox{\boldmath$\omega$}| )$ and the dissipation of turbulent energy. Wallace {\textit{et al.}}\  \cite{wal1992} performed an elaborate experimental study of the helicity density in a turbulent boundary-layer, a two-stream mixing-layer, and in grid-flow turbulence. They found that there is a tendency for the instantaneous velocity and vorticity to align in the shear flows, but concluded that there is little relationship between the small instantaneous dissipation and large helicity density except in the shear flows. Their results support the numerical results obtained by Rogers and Moin.

In general, the second-order correlation tensor of the velocity for
homogeneous and isotropic but non-mirrorsymmetric turbulence is expressed
in terms of the energy (pure scalar) and helicity (pseudoscalar)
profiles. Note that the helicity-related part never appears in the
mirrorsymmetric case. It has been argued from the symmetry of the
Reynolds-stress tensor that helicity itself cannot contribute to the
Reynolds stress \cite{kra1974}. It has also been pointed out that
the presence of turbulent helicity density alone is insufficient and
some other factors breaking the symmetry, such as the compressibility
\cite{moi1983,chk1988,kho1991,kit1994}, anisotropy \cite{fri1987},
mean flow, etc.\ are indispensable for the large-scale vortical flow
generation. In this context, it is important to note that Gvaramadze
{\textit{et al.}}\ \cite{gva1989} showed that even in incompressible
turbulence, turbulent helicity may contribute to the generation of
large-scale vortices through the coupling with the mean flow. Also
Chkhetiany {\textit{et al.}}\ \cite{chk1994} showed possibility of a
spontaneous generation of vortical structures in homogeneous turbulent
shear with helicity.

	Assuming the general form of the correlation functions for homogeneous isotropic and non-mirrorsymmetric turbulence profiles as the basic or lowest-order field in the framework of a closure theory for inhomogeneous turbulence, Yokoi and Yoshizawa \cite{yok1993} obtained an expression for the Reynolds stress from the fundamental fluid equations. In this expression, the gradient of the turbulent helicity enters the Reynolds stress as a higher-order effect representing the mean-field inhomogeneities. In their formulation with a derivative expansion [see Eqs.~(\ref{eq:two-scale_diff_exp}) and (\ref{eq:diff_exp}) in \S~\ref{sec:3B}] with respect to the large-scale inhomogeneities, the gradient of helicity appears as the coupling coefficient for the mean vorticity in non-mirrorsymmetric turbulence.

	At the same time, it had been well recognized that the usual turbulence model with the eddy-viscosity expression for the Reynolds stress completely fails when it is applied to a turbulent swirling flow \cite{kob1987}. With the usual eddy-viscosity model, the dent or decelerated profile of the mean axial velocity near the center axis, imposed at the inlet, cannot be sustained and will decay rapidly and immediately to turn to the usual flat profile of the non-swirling pipe flow. The turbulent or eddy viscosity is too strong to produce an inhomogeneity in the axial velocity profile. Yokoi and Yoshizawa \cite{yok1993} applied their turbulence model with the helicity effect implemented into the Reynolds stress to a turbulent swirling flow and succeeded in reproducing the dent profile in the downstream region found experimentally \cite{kit1991,ste1995}. In this sense, the inhomogeneous helicity effect has been confirmed at the level of turbulence model simulations or closure calculations.
	
	As has been referred to above, the theoretical derivation of the Reynolds-stress expression is very general and straightforward. It was based not on the heuristic assumptions for Reynolds stress modeling but on the generic expression for the turbulence fields that are isotropic and non-mirrorsymmetric in the lowest order correlation. However, the closure scheme itself contains several approximations \cite{yos1984,yok2013}. It is necessary to study carefully the inhomogeneous helicity effect in DNSs. One of the most straightforward tests is to check the model expression for the Reynolds stress and compare it to the DNS result for the Reynolds stress. Since the transport coefficients in turbulence models are expressed in terms of turbulent statistical quantities such as the turbulent energy, its dissipation rate, the turbulent helicity, etc., we have to calculate the spatiotemporal evolution of  the statistical quantities using DNS data. In the present work, we perform DNSs of inhomogeneous helical turbulence with the simplest possible flow geometry, and validate the turbulence model expression based on the theoretical investigation.

	The organization of this paper is as follows. After presenting the fundamental equations in \S~\ref{sec:2}, we summarize the helicity effects in inhomogeneous turbulence with special reference to the symmetry of Reynolds stress and its modeling in \S~\ref{sec:3}. In \S~\ref{sec:4}, the set-up of the numerical simulation and its results are presented. In \S~\ref{sec:5}, the helicity effect on turbulent momentum transport is discussed with a special reference to the vortex dynamo. Conclusions are given in \S~\ref{sec:6}. Details of the turbulence model with helicity and its application to a turbulent swirling flow, comparison with previous notions including the so-called $\Lambda$ effect and the Anisotropic Kinetic Alpha (AKA) effect are given in Appendices.

\section{Fundamental equations\label{sec:2}}
	We consider an incompressible fluid in a rotating system. The velocity ${\bf{u}}$ obeys the incompressible Navier--Stokes equation
\begin{equation}
	\frac{\partial{\bf{u}}}{\partial t}
	+ ({\bf{u}} \cdot \nabla) {\bf{u}}
	= - \nabla p
	+ {\bf{u}} \times 2 \mbox{\boldmath$\omega$}_{\rm{F}}
	+ \nu \nabla^2 {\bf{u}}
	+ {\bf{f}}_{\rm{e}}
	\label{eq:NS_eq}
\end{equation}
and the solenoidal condition
\begin{equation}
	\nabla \cdot {\bf{u}} = 0,
	\label{eq:solenoidal_u}
\end{equation}
where $p$ is the pressure divided by fluid density with the centrifugal force included, $\nu$ the kinematic viscosity, $\mbox{\boldmath$\omega$}_{\rm{F}}$ the angular velocity of the system, and ${\bf{f}}_{\rm{e}}$ the external forcing which satisfies the solenoidal conditions.

	Taking a curl operation to Eqs.~(\ref{eq:NS_eq}) and (\ref{eq:solenoidal_u}), we have the equations of vorticity $\mbox{\boldmath$\omega$} (= \nabla \times {\bf{u}})$ as
\begin{equation}
	\frac{\partial \mbox{\boldmath$\omega$}}{\partial t}
	= \nabla \times \left[ {
		{\bf{u}} \times \left( {
			\mbox{\boldmath$\omega$} + 2 \mbox{\boldmath$\omega$}_{\rm{F}}
		} \right)
	} \right]
	+ \nu \nabla^2 \mbox{\boldmath$\omega$}
	+ \nabla \times {\bf{f}}_{\rm{e}}
	\label{eq:omega_eq}
\end{equation}
and
\begin{equation}
	\nabla \cdot \mbox{\boldmath$\omega$} = 0.
	\label{eq:solenoidal_omega}
\end{equation}

	We divide a flow quantity $f$ into the mean part $\langle {f} \rangle$ and fluctuation around it, $f'$, as
\begin{subequations}
\begin{equation}
	f = F + f',\; F = \langle {f} \rangle
	\label{rey_decomp}
\end{equation}
with
\begin{equation}
	f = ({\bf{u}}, p,\mbox{\boldmath$\omega$}),\;
	F = ({\bf{U}}, P,\mbox{\boldmath$\Omega$}),\;
	f' = ({\bf{u}}', p',\mbox{\boldmath$\omega$}').
	\label{rey_decomp_flds}
\end{equation}
\end{subequations}
Substituting Eq.~(\ref{rey_decomp}) into Eqs.~(\ref{eq:NS_eq})-(\ref{eq:solenoidal_omega}), we obtain the mean-field equations as
\begin{equation}
	\frac{\partial{\bf{U}}}{\partial t}
	+ ({\bf{U}} \cdot \nabla) {\bf{U}}
	= - \nabla P
	+ {\bf{U}} \times 2 \mbox{\boldmath$\omega$}_{\rm{F}}
	- \nabla \cdot \mbox{\boldmath${\cal{R}}$}
	+ \nu \nabla^2 {\bf{U}},
	\label{eq:mean_vel_eq}
\end{equation}
\begin{equation}
	\frac{\partial \mbox{\boldmath$\Omega$}}{\partial t}
	= \nabla \times \left[ {
		{\bf{U}} \times \left( {
			\mbox{\boldmath$\Omega$} + 2 \mbox{\boldmath$\omega$}_{\rm{F}}
		} \right)
	} \right]
	+ \nabla \times {\bf{V}}_{\rm{M}}
	+ \nu \nabla^2 \mbox{\boldmath$\Omega$},
	\label{eq:mean_vor_eq}
\end{equation}
\begin{equation}
	\nabla \cdot {\bf{U}}
	= \nabla \cdot \mbox{\boldmath$\Omega$} 
	= 0,
	\label{eq:solenoidal_mean}
\end{equation}
where $\mbox{\boldmath${\cal{R}}$} = \{ {{\cal{R}}^{ij}} \}$ and ${\bf{V}}_{\rm{M}}$ are the Reynolds stress and the turbulent Vortex-Motive or Pondero-Motive Force (VMF or PMF), which are defined by
\begin{equation}
	{\cal{R}}^{ij}
	= \left\langle {u'{}^i u'{}^j} \right\rangle,
	\label{eq:rey_strss_def_2}
\end{equation}
\begin{equation}
	{\bf{V}}_{\rm{M}} = \langle {{\bf{u}}' \times \mbox{\boldmath$\omega$}'} \rangle,
	\label{eq:vmf_def_2}
\end{equation}
respectively. If we compare Eqs.~(\ref{eq:mean_vel_eq}) and (\ref{eq:mean_vor_eq}) with Eqs.~(\ref{eq:NS_eq}) and (\ref{eq:omega_eq}), we see the Reynolds stress and the VMF are the sole quantities that represent the effects of turbulence in the mean-field equations. It should be noted that in this work we adopt an external forcing that does not directly produce any mean flow ($\langle {{\bf{f}}_{\rm{e}}} \rangle = 0$).

\section{Helicity effect\label{sec:3}}
\subsection{Symmetry of the Reynolds stress\label{sec:3A}}
	As was mentioned in \S~\ref{sec:1}, the presence of turbulent kinetic helicity alone is not sufficient for the helicity effect to appear in the mean momentum transport. As we will see later in Eq.~(\ref{eq:Rey_strss_model}), the inhomogeneity of the turbulent helicity ($\nabla H$) is a key ingredient. This point is easily understood if we consider the symmetry properties of the Reynolds stress tensor
\begin{equation}	
	{\cal{R}}^{ij}({\bf{x}},t)
	= \left\langle {u'{}^i({\bf{x}},t) u'{}^j({\bf{x}},t)} \right\rangle
	\label{eq:Rey_strss_def}
\end{equation}
(${\bf{u}}'$: velocity fluctuation). Helicity is a quantity which represents the breakage of mirror- or reflectional symmetry. A reflection with respect to a plane is equivalent to the combination of a pure (proper) rotation around the axis perpendicular to the plane and an inversion or parity transformation. Since proper rotations never change the mirrorsymmetry-related property of a vector, we can express the symmetry property of reflections in terms of that of inversions. (Note that the determinant of the transformation is always $+1$ for all proper rotations whereas the counterparts are $-1$ for reflections and inversions.) The velocity is a polar vector and has odd parity under inversion. Namely, with a reversal of the coordinate system: $x^i \longmapsto \tilde{x}^i = - x^i$ (a tilde denotes a quantity in the reversal frame), the velocity  reverses its sign as $u^i({\bf{x}},t) \longmapsto \tilde{u}^i(\tilde{\bf{x}}, t) = - u^i({\bf{x}},t)$. As a consequence, the Reynolds stress transforms under inversion as
\begin{eqnarray}
	&&{\cal{R}}^{ij}({\bf{x}},t)
	\nonumber\\
	&&\longmapsto \tilde{\cal{R}}^{ij}(\tilde{\bf{x}}, t) 
	= \langle {
		\tilde{u}'{}^i(\tilde{\bf{x}},t) \tilde{u}'{}^j(\tilde{\bf{x}},t)
	} \rangle
	\nonumber\\
	&&\hspace{20pt} =  \langle {
		[-u'{}^i({\bf{x}},t)] [-u'{}^j({\bf{x}},t)]
	} \rangle
	= \langle {
		u'{}^i({\bf{x}},t) u'{}^j({\bf{x}},t)
	} \rangle
	\nonumber\\
	&&\hspace{20pt}= {\cal{R}}^{ij}({\bf{x}},t).
	\label{eq:Rey_strss_parity}
\end{eqnarray}
Namely, the Reynolds stress is symmetric with respect to the inversion of the coordinate system and must have even parity.

	The mean vorticity $\mbox{\boldmath$\Omega$}$ ($= \nabla \times {\bf{U}}$, ${\bf{U}}$: mean velocity) is an axial- or pseudo-vector which does not change its sign (symmetric) under the inversion:
\begin{eqnarray}
	&&\Omega^i({\bf{x}},t)
	\nonumber\\
	&&\longmapsto \tilde{\Omega}^i(\tilde{\bf{x}},t)
	= \tilde{\epsilon}^{ijk} 
		\frac{\partial \tilde{U}^k(\tilde{\bf{x}},t)}{\partial \tilde{x}^j}
	\nonumber\\
	&&\hspace{20pt}= \epsilon^{ijk} \frac{\partial (-U)^k({\bf{x}},t)}{\partial (-x)^j}
	= \epsilon^{ijk} \frac{\partial U^k({\bf{x}},t)}{\partial x^j}
	\nonumber\\
	&&\hspace{20pt}= \Omega^i ({\bf{x}},t)
	\label{eq:Omega_parity}
\end{eqnarray}	
(Note that the alternate tensor has even parity: $\epsilon^{ijk} \longmapsto \tilde{\epsilon}^{ijk} = \epsilon^{ijk}$). On the other hand, the turbulent helicity $H (= \langle {{\bf{u}}' \cdot \nabla \times {\bf{u}}'} \rangle)$ is a pseudoscalar which changes its sign (antisymmetric) under the inversion:
\begin{eqnarray}
	&&H({\bf{x}},t)
	\nonumber\\ 
	&&\longmapsto \tilde{H}(\tilde{\bf{x}},t)
	= \left\langle {
		\tilde{u}'{}^i(\tilde{\bf{x}},t)
		\tilde{\epsilon}^{ijk} 
		\frac{\partial {\tilde{u}}'{}^k(\tilde{\bf{x}},t)}{\partial \tilde{x}^j}
	} \right\rangle
	\nonumber\\
	&&\hspace{10pt}= \left\langle {
		-u'{}^i({\bf{x}},t)
		\epsilon^{ijk} \frac{\partial (-u'{}^k)({\bf{x}},t)}{\partial (-x)^j}
	} \right\rangle
	\nonumber\\
	&&\hspace{10pt}= - \left\langle {
		u'{}^i({\bf{x}},t)
		\epsilon^{ijk} \frac{\partial u'{}^k({\bf{x}},t)}{\partial x^j}
	} \right\rangle
	= - H({\bf{x}},t).
	\label{eq:H_parity}
\end{eqnarray}	
From Eq.~(\ref{eq:H_parity}) we infer an important point. In a mirrorsymmetric system, by definition, all the statistical quantities are symmetric under inversion (or reflection) as
\begin{equation}
	F({\bf{x}},t) 
	\longmapsto \hat{F}({\bf{x}},t) = F({\bf{x}},t).
	\label{eq:mirrorsym_system}
\end{equation}
On the other hand, any pseudoscalar changes its sign under the inversion (or reflection) as
\begin{equation}
	F({\bf{x}},t) \longmapsto \hat{F}({\bf{x}},t) = - F({\bf{x}},t).
	\label{eq:pseudo_scalar}
\end{equation}
It follows from Eqs.~(\ref{eq:mirrorsym_system}) and (\ref{eq:pseudo_scalar}) that any pseudoscalar statistical quantity in a mirrorsymmetric system should satisfy
\begin{equation}
	F({\bf{x}},t) = - F({\bf{x}},t)
	\label{eq:vanishing_peudoscalar_1}
\end{equation}
or equivalently,
\begin{equation}
	F({\bf{x}},t) = 0.
	\label{eq:vanishing_peudoscalar_2}
\end{equation}
Hence, any pseudoscalar statistical quantities should vanish in a mirrorsymmetric system. In other words, a finite pseudoscalar indicates a broken mirrorsymmetry in the system. In this sense, a pseudoscalar statistical quantity can serve itself as a measure of the breakage of mirrorsymmetry. Since the helicity $\int_V{{\bf{u}} \cdot \mbox{\boldmath$\omega$}}dV$, as well as the kinetic energy $\int_V {{\bf{u}}^2} dV$, is an inviscid invariant of the hydrodynamic equation, its local turbulent density, the turbulent helicity $H \equiv \langle {{\bf{u}}' \cdot \mbox{\boldmath$\omega$}'} \rangle$ [$\mbox{\boldmath$\omega$}' (= \nabla \times {\bf{u}}')$: vorticity fluctuation], is an important statistical quantity that represents structural properties of the turbulence.

A positive (negative) sign of local helicity represents right-handed
(left-handed) ``twistedness'' of the turbulence. The sign of $H$ is directly
connected to the structural properties of turbulence. However, as
explained below, $H$ itself cannot enter the Reynolds stress expression. We saw
in Eq.~(\ref{eq:Omega_parity}) that the mean vorticity or rotation vector
has even parity. So, the coupling coefficient for the mean vorticity or
rotation should have even parity in order to attain the even parity for
the Reynolds stress [Eq.~(\ref{eq:Rey_strss_parity})]. This suggests that
the turbulent helicity with odd parity [Eq.~(\ref{eq:H_parity})] itself
cannot enter the expression for the Reynolds stress as the coupling
coefficient for the mean vorticity or the rotation velocity. This point is
reflected later in the generic mathematical expression of the correlation
in non-mirrorsymmetric isotropic turbulence, Eq.~(\ref{eq:iso_nonmirror}).

\subsection{Helicity effect in the Reynolds stress\label{sec:3B}}
	Using the Two-Scale Direct-Interaction Approximation (TSDIA), a closure theory for inhomogeneous turbulence \cite{yos1984}, Yokoi and Yoshizawa \cite{yok1993} explored the effects of helicity in inhomogeneous turbulence. The TSDIA is a combination of the multiple-scale analysis and the DIA, an elaborated renormalized perturbation method in ${\bf{k}}$, or wavenumber space, for homogeneous turbulence at high Reynolds number. 
	
	In this analysis, we introduce two scales for space and time variables with a scale parameter $\delta$ as
\begin{equation}
	\mbox{\boldmath$\xi$} = {\bf{x}},\; {\bf{X}} = \delta {\bf{x}};\;\;
	\tau = t,\; T = \delta t,
	\label{eq:two-scale_variables}
\end{equation}
where ($\mbox{\boldmath$\xi$}, \tau$) and (${\bf{X}}, T$) are fast and slow variables, respectively. With a small $\delta$, the slow variables (${\bf{X}}, T$) are suitable for representing slow variations of fields since they vary only when ${\bf{x}}$ and $t$ vary strongly. Under Eq.~(\ref{eq:two-scale_variables}), a field quantity is divided into the mean $F$ and the fluctuation $f'$ as
\begin{equation}
	f = F({\bf{X}};T) + f'(\mbox{\boldmath$\xi$},{\bf{X}}; \tau,T),
	\label{two-scale_fields}
\end{equation}
which represents the properties that the mean field slowly changes with slow variables and fluctuation field depends both on fast and slow variables.
Also under Eq.~(\ref{eq:two-scale_variables}), we have  
\begin{equation}
	\nabla = \nabla_{\small{\mbox{\boldmath$\xi$}}} + \delta \nabla_{\bf{X}},\;\;
	\frac{\partial}{\partial t} = \frac{\partial}{\partial \tau} + \delta \frac{\partial}{\partial T},
	\label{eq:two-scale_diff_exp}
\end{equation}
where $\nabla_{{\mbox{\boldmath$\xi$}}}^i = (\partial / \partial \xi^i)$ and $\nabla_{\bf{X}}^i = (\partial / \partial X^i)$. We see from Eq.~(\ref{eq:two-scale_diff_exp}) that a derivative with respect to the slow variables gives an $O(\delta)$ contribution (derivative expansion) \cite{nay1973}. 

	The turbulence correlations such as the Reynolds stress are calculated with the aid of a perturbation expansion:
\begin{equation}
	f'({\bf{k}},{\bf{X}}; \tau,T) 
	= \sum_{n=0}^{\infty} \delta^n  f'_n({\bf{k}},{\bf{X}}; \tau,T).
	\label{eq:diff_exp}
\end{equation}
The scale parameter $\delta$ is associated with the inhomogeneity of the large-scale fields.

	The $O(\delta^0)$ fields correspond to those in homogeneous turbulence. We also expand the lowest- and higher-order fields with respect to the rotation vector $\mbox{\boldmath$\omega$}_{\rm{F}}$ as
\begin{eqnarray}
	{\bf{u}}'_0({\bf{k}},{\bf{X}}; \tau,T)
	&=& {\bf{u}}'_{\rm{B}}({\bf{k}},{\bf{X}}; \tau,T)
	\nonumber\\
	&& + \sum_{m=1}^{\infty}  |\mbox{\boldmath$\omega$}_{\rm{F}}|^m 
		{\bf{u}}'_{0m}({\bf{k}},{\bf{X}}; \tau,T),
	\label{eq:u0_expn}
\end{eqnarray}
\begin{equation}
	{\bf{u}}'_n({\bf{k}},{\bf{X}}; \tau,T)
	= \sum_{m=0}^{\infty}  |\mbox{\boldmath$\omega$}_{\rm{F}}|^m 
		{\bf{u}}'_{nm}({\bf{k}},{\bf{X}}; \tau,T)\;\;
	(n\ge 1).
	\label{eq:un_expn}
\end{equation}
	
	The basic or lowest-order field ${\bf{u}}_{\rm{B}}$ corresponds to a homogeneous isotropic field, and the effects of inhomogeneity and anisotropy are systematically incorporated in the higher-order field calculations using the spectral and response functions.

	As for the statistical properties of the basic fields, we assume 
\begin{eqnarray}
	&&{\left\langle {
		u'_{\rm{B}}{}^\alpha({\bf{k}},{\bf{X}};\tau,T)
		u'_{\rm{B}}{}^\beta({\bf{k}}',{\bf{X}};\tau',T)
	} \right\rangle}/{\delta({\bf{k}} + {\bf{k}}')}
	\nonumber\\
	&&\hspace{10pt} 
	= D^{\alpha\beta}({\bf{k}}) Q_{\rm{B}}(k,{\bf{X}};\tau,\tau',T)
	\nonumber\\
	&&\hspace{20pt} + \frac{i}{2} \frac{k^a}{k^2} \epsilon^{\alpha\beta a} 
		H_{\rm{B}}(k,{\bf{X}};\tau',T),
  \label{eq:iso_nonmirror}
\end{eqnarray}
\begin{equation}
	\left\langle {G^{\alpha\beta}({\bf{k}},{\bf{X}};\tau,\tau',T)} \right\rangle
	= D^{\alpha\beta}({\bf{k}}) G(k,{\bf{X}};\tau,\tau',T),
	\label{eq:G_iso}
\end{equation}
where $Q_{\rm{B}}$ and $H_{\rm{B}}$ are the spectral density functions of the turbulent energy and helicity, respectively:
\begin{equation}
	\frac{1}{2} \left\langle {{\bf{u}}'_{\rm{B}}{}^2} \right\rangle
	= \int d{\bf{k}} Q_{\rm{B}}(k; \tau, \tau),
	\label{eq:Q_B_spectrum}
\end{equation}
\begin{equation}
	\left\langle {{\bf{u}}'_{\rm{B}} \cdot \mbox{\boldmath$\omega$}'_{\rm{B}}} \right\rangle
	= \int d{\bf{k}} H_{\rm{B}}(k, \tau, \tau)
	\label{eq:H_B_spectrum}
\end{equation}
and $D^{\alpha\beta}({\bf{k}}) (= \delta^{\alpha\beta}
- k^\alpha k^\beta/k^2)$ is the solenoidal projection
operator. It should be noticed that $Q_{\rm{B}}$ and $H_{\rm{B}}$
in Eq.~(\ref{eq:iso_nonmirror}) are normalized to satisfy
Eqs.~(\ref{eq:Q_B_spectrum}) and (\ref{eq:H_B_spectrum}). Equations~(\ref{eq:iso_nonmirror}) and
(\ref{eq:G_iso}) are the most general expressions for homogeneous,
isotropic and non-mirrorsymmetric turbulence \cite{bat1953,mon1975}. We
should note that these assumptions only apply to the basic or lowest-order
fields of turbulence. The turbulent fields considered in this formulation
are inhomogeneous and anisotropic; these effects enter through the
higher-order fields.

	The Reynolds stress is calculated by
\begin{eqnarray}
	\left\langle {u'{}^\alpha u'{}^\beta} \right\rangle
	&=& \left\langle {u'_{\rm{B}}{}^\alpha u'_{\rm{B}}{}^\beta} \right\rangle
	+ \left\langle {u'_{\rm{B}}{}^\alpha u'_{01}{}^\beta} \right\rangle
	+ \left\langle {u'_{01}{}^\alpha u'_{\rm{B}}{}^\beta} \right\rangle
	+ \cdots
	\nonumber\\
	&&+ \left\langle {u'_{\rm{B}}{}^\alpha u'_{10}{}^\beta} \right\rangle
	+ \left\langle {u'_{10}{}^\alpha u'_{\rm{B}}{}^\beta} \right\rangle
	+ \cdots.
	\label{eq:Rey_strss_expn}
\end{eqnarray}
It was shown by the analysis up to $O(\delta^1 |\mbox{\boldmath$\omega$}_{\rm{F}}|^1)$ that the Reynolds stress is expressed as \cite{yok1993}
\begin{eqnarray}
	\lefteqn{
	\left\langle {u'{}^\alpha u'{}^\beta} \right\rangle_{\rm{D}}
	= - \nu_{\rm{T}} {\cal{S}}^{\alpha\beta}
	}\nonumber\\
	&&\hspace{10pt} + \left[ {
		\Gamma^\alpha \left( {
			\Omega^\beta + 2 \omega_{\rm{F}}^\beta
		} \right)
		+ \Gamma^\beta \left( {
			\Omega^\alpha + 2 \omega_{\rm{F}}^\alpha
		} \right)
	} \right]_{\rm{D}},
	\label{eq:Rey_strss_exprssn}
\end{eqnarray}
where ${\rm{D}}$ denotes the deviatoric or traceless part of tensor as ${\cal{A}}^{\alpha\beta}_{\rm{D}} = {\cal{A}}^{\alpha\beta} - (1/3) {\cal{A}}^{aa} \delta^{\alpha\beta}$, and ${\cal{S}}$ is the mean velocity strain defined by
\begin{equation}
	{\cal{S}}^{\alpha\beta}
	= \frac{\partial U^\alpha}{\partial x^\beta}
	+ \frac{\partial U^\beta}{\partial x^\alpha}
	- \frac{2}{3} \nabla \cdot {\bf{U}} \delta^{\alpha\beta}.
	\label{eq:mean_vel_strn_def}
\end{equation}
In Eq.~(\ref{eq:Rey_strss_exprssn}), the mean velocity-strain- and the mean vorticity and angular velocity-related coefficients, $\nu_{\rm{T}}$ and $\mbox{\boldmath$\Gamma$}$, are given by
\begin{equation}
	\nu_{\rm{T}} 
	= \frac{7}{15} \int {\rm{d}}{\bf{k}} 
		\int_{-\infty}^{t} \!\!\!d\tau_1\ G(k;\tau,\tau_1) {Q}(k;\tau,\tau_1),
	\label{eq:nu_T_exprssn}
\end{equation}
\begin{equation}
	\mbox{\boldmath$\Gamma$} 
	= \frac{1}{30} \int k^{-2} {\rm{d}}{\bf{k}} 
	\int_{-\infty}^{t} \!\!\!d\tau_1\ G(k;\tau,\tau_1) \nabla {H}(k;\tau,\tau_1).
	\label{eq:Gamma_exprssn}
\end{equation}

	The first term of Eq.~(\ref{eq:Rey_strss_exprssn}) corresponds to the usual eddy-viscosity representation of the Reynolds stress. The second term is the correction to the eddy-viscosity representation due to the mean vortical or rotational motion.

	Equations~(\ref{eq:nu_T_exprssn}) and (\ref{eq:Gamma_exprssn}) show that we can express the turbulent transport coefficients if we know the propagators of turbulent field such as the spectral functions of energy and helicity, ${{Q}}(k;\tau,\tau')$ and ${{H}}(k;\tau,\tau')$, and the response function $G(k;\tau,\tau')$, which represent how turbulence is distributed in scales and how much the present state is affected by the past, respectively.

	However, for practical purpose, the theoretical expressions for the transport coefficients $\nu_{\rm{T}}$ [Eq.~(\ref{eq:nu_T_exprssn})] and $\mbox{\boldmath$\Gamma$}$ [Eq.~(\ref{eq:Gamma_exprssn})] in terms of the time and spectral integrals of the propagators are too much complicated. We need to reduce them into a more tractable form. In the simplest case, if the time integral of the response function can be separated from the spectral integral of the energy in Eq.~(\ref{eq:nu_T_exprssn}), the time integral of the response function just gives a time scale of turbulence, $\tau$:
\begin{equation}
	\tau \simeq \int_{-\infty}^{t} \!\!\!d\tau_1\ G(k;\tau,\tau_1). 
	\label{eq:time_G}
\end{equation}
Then, Eq.~(\ref{eq:nu_T_exprssn}) is reduced to the turbulence time scale multiplied by the turbulent energy $K$. Namely, the mixing-length expression for the turbulent viscosity:
\begin{equation}
	\nu_{\rm{T}} \sim \tau K \sim \tau u^2 \sim \ell u.
	\label{nu_T_mxng_lngth}
\end{equation}
In other words, Eq.~(\ref{eq:nu_T_exprssn}) is a natural generalization of the simplest mixing-length expression for the turbulent viscosity.

\subsection{Helicity turbulence model\label{sec:3C}}
	The Reynolds-stress expression (\ref{eq:Rey_strss_exprssn}) with the spectral expressions for the transport coefficients [Eqs.~(\ref{eq:nu_T_exprssn}) and (\ref{eq:Gamma_exprssn})] is too heavy for practical uses in the astro/geophysical applications. In order to construct simple expressions for the transport coefficients more generic than the mixing-length one, we use one-point turbulence statistical quantities which represent the statistical properties of turbulence. We choose the turbulent energy $K$, its dissipation rate $\varepsilon$, and the turbulent helicity $H$, defined by
\begin{equation}
	K = \frac{1}{2} \left\langle {{\bf{u}}'{}^2} \right\rangle,
	\label{eq:K_def}
\end{equation}
\begin{equation}
	\varepsilon = \nu \left\langle {
		\frac{\partial u'{}^a}{\partial x^b}
		\frac{\partial u'{}^a}{\partial x^b}
	} \right\rangle,
	\label{eq:eps_def}
\end{equation}
\begin{equation}
	H = \left\langle {
		{\bf{u}}' \cdot \mbox{\boldmath$\omega$}'
	} \right\rangle.
	\label{eq:H_def}
\end{equation}
On the basis of  the analytical expression Eq.~(\ref{eq:Rey_strss_exprssn}), the Reynolds stress is modeled as \cite{yok1993}
\begin{eqnarray}
	&&\left\langle { u'{}^\alpha u'{}^\beta
	} \right\rangle_{\rm{D}}
	= - \nu_{\rm{T}} {\cal{S}}^{\alpha\beta}
	\nonumber\\
	&& \hspace{10pt} + \eta \left[ {
		\frac{\partial H}{\partial x^\alpha} \left( {
        \Omega^\beta + 2\omega_{\rm{F}}^\beta
        } \right)
		+ \frac{\partial H}{\partial x^\beta} \left( {
        \Omega^\alpha + 2\omega_{\rm{F}}^\alpha
        } \right)
	} \right]_{\rm{D}}
	\label{eq:Rey_strss_model}
\end{eqnarray}
with the transport coefficients expressed in terms of the above turbulence statistical quantities [Eqs.~(\ref{eq:K_def})-(\ref{eq:H_def})]:
\begin{equation}
	\nu_{\rm{T}} 
	= C_\nu \tau K
	= C_\nu \frac{K}{\varepsilon} K,
	\label{eq:nu_T_K_eps}
\end{equation}
\begin{equation}
	\eta 
	= C_\eta \tau \ell^2
	= C_\eta \frac{K}{\varepsilon} \frac{K^3}{\varepsilon^2},
	\label{eq:eta_K_eps}
\end{equation}
where $C_\nu$ and $C_\eta$ are model constants, whose values are to be optimized through the applications of the turbulence model to several flows \cite{lau1972}. An application of the helicity model to a turbulent swirling flow \cite{yok1993} suggests
\begin{equation}
	C_\nu = 0.09,\;\;
	C_\eta = 0.003.
	\label{eq:model_consts_hel}
\end{equation}
Note that, in Eqs.~(\ref{eq:nu_T_K_eps}) and (\ref{eq:eta_K_eps}), $\tau = K / \varepsilon$ and $\ell = K^{3/2} / \varepsilon$ are the time and length scales of turbulence, respectively. Equation~(\ref{eq:eta_K_eps}) corresponds to the modeling of the mean vorticity- and angular-velocity-related coefficient $\mbox{\boldmath$\Gamma$}$ [Eq.~(\ref{eq:Gamma_exprssn})] as
\begin{equation}
	\mbox{\boldmath$\Gamma$}
	= \eta \nabla H 
	= C_\eta \tau \ell^2 \nabla H
	= C_\eta \frac{K}{\varepsilon} \frac{K^3}{\varepsilon^2} \nabla H.
	\label{eq:Gamma_model}
\end{equation}

	As is seen in Eqs.~(\ref{eq:nu_T_K_eps}), (\ref{eq:eta_K_eps}), and (\ref{eq:Gamma_model}), the transport coefficients in the Reynolds stress, $\nu_{\rm{T}}$, $\eta$, and $\mbox{\boldmath$\Gamma$}$, are expressed in terms of the turbulent statistical quantities, $K$, $\varepsilon$, and $H$. In order to close or construct a self-consistent turbulence model, we consider the transport equations of $K$, $\varepsilon$, and $H$. Details of the present helicity model with the transport equations of $K$, $\varepsilon$, and $H$ is presented in Appendix \ref{sec:appA1}.
	
	The helicity turbulence model was applied to a turbulent swirling pipe flow \cite{yok1993}. It was numerically shown that the model successfully reproduces basic behaviors of the turbulent swirling flow: (i) the stationary dent profile of the mean axial velocity in the central axis region; (ii) the radial profile of the mean circumferential velocity; (iii) the exponential decay of the swirl intensity defined by the axial flux of the mean angular momentum along the axial direction \cite{kit1991,ste1995}. These behavior could not be reproduced by the standard $K - \varepsilon$ model with the eddy viscosity \cite{kob1987}. In this sense, validity of the Reynolds-stress expression [Eqs.~(\ref{eq:Rey_strss_exprssn}) and (\ref{eq:Rey_strss_model})] is confirmed at the turbulence or closure model simulation level. Details of the application of the helicity turbulence model to a swirling pipe flow are presented in Appendix~\ref{sec:appA2}.

\section{Numerical simulations\label{sec:4}}
	It is necessary to study carefully the inhomogeneous helicity effect in the turbulent momentum transport using Direct Numerical Simulations (DNSs). For this purpose, we check the validity of the expression for the Reynolds stress [Eq.~(\ref{eq:Rey_strss_model})] using DNSs of a turbulent flow with inhomogeneous helicity.

\subsection{Set-up\label{sec:4A}}
\begin{figure}[b!]
\includegraphics[width=.40\textwidth]{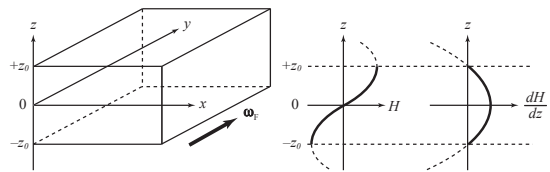}
\caption{Set-up of the turbulence with rotation $\mbox{\boldmath$\omega$}_{\rm{F}}$ (left) and schematic spatial profiles of turbulent helicity $H$ ($= \langle {{\bf{u}}' \cdot \mbox{\boldmath$\omega$}'} \rangle$) given by Eq.~(\ref{eq:H_profile}) (center) and its derivative $dH/dz$ (right). The helicity inhomogeneity is generated by an external forcing.}
\label{fig:setup} 
\end{figure}

	In the present work, we adopt a set-up that is suitable for examining the deviatoric part of the Reynolds-stress expression [Eqs.~(\ref{eq:Rey_strss_exprssn}) and (\ref{eq:Rey_strss_model})]. Let us consider helical turbulence in a box with imposed rotation $\mbox{\boldmath$\omega$}_{\rm{F}}$ depicted in Fig.~\ref{fig:setup}. The axis of rotation $\mbox{\boldmath$\omega$}_{\rm{F}}$ is aligned with the $y$ axis as
\begin{equation}
	\mbox{\boldmath$\omega$}_{\rm{F}}
	= (\omega_{\rm{F}}^x, \omega_{\rm{F}}^y, \omega_{\rm{F}}^z)
	= (0, \omega_{\rm{F}}, 0).
	\label{eq:omega_F_setup}
\end{equation}

	The inhomogeneous helicity is sustained by an external forcing, leading to a spatial distribution of turbulent helicity schematically expressed as
\begin{equation}
	H(z) = - \frac{1}{2} H_0 z (z^2 - 3z_0^2),
	\label{eq:H_profile}
\end{equation}
where $H_0$ is the peak magnitude of the turbulent helicity at positions $z = \pm z_0$.

	In the simulations, we use helically forced turbulence where the degree of helicity is modulated in the $z$ direction in a periodic fashion. The wavenumber of the forcing is $\kf$ and that of the box is $k_1$, which is also the wavenumber of the helicity modulation in the $z$ direction. 

	We consider averages over $x$, $y$, and $t$ over an interval during which the system is statistically steady. We denote these averages by $\langle \cdots \rangle$. In particular, we consider the Reynolds stress tensor component ${\cal{R}}^{yz}=\langle{u'{}^y u'{}^z}\rangle$, the mean flow $\langle {{\bf{u}}} \rangle = {\bf{U}} (z,t)$, and the helicity density $H = \langle {{\bf{u}}' \cdot \mbox{\boldmath$\omega$}'} \rangle$ with $\mbox{\boldmath$\omega$}' = \nabla \times {\bf{u}}'$ being the vorticity fluctuation.
In all cases, the helicity of the mean flow, ${\bf{U} \cdot \mbox{\boldmath$\Omega$}}$ ($\mbox{\boldmath$\Omega$} = \nabla \times {\bf{U}}$), is negligible.

We apply rotation in the $y$ direction as Eq.~(\ref{eq:omega_F_setup}). At
the initial stage, we have no large-scale flow ${\bf{U}}$. It follows
from Eq.~(\ref{eq:Rey_strss_model}) that, at the early stage of flow
evolution, the $y$-$z$ component of the Reynolds stress may be given by
\begin{equation}
	\langle {u'{}^y u'{}^z} \rangle
	= \eta 2 \omega_{\rm{F}}^y \frac{\partial H}{\partial z}.
	\label{eq:Rey_strss_num_early}
\end{equation}
Once the large-scale flow is generated, but the mean relative vorticity is still smaller than the rotation ($|\mbox{\boldmath$\Omega$}| \ll |2 \mbox{\boldmath$\omega$}_{\rm{F}}|$), the $y$-$z$ component of the Reynolds stress is written as
\begin{equation}
	\langle {u'{}^y u'{}^z} \rangle
	= -\nuT \frac{\partial U^y}{\partial z}
	+ \eta 2 \omega_{\rm{F}}^y \frac{\partial H}{\partial z}.
	\label{eq:Rey_strss_num_dev}
\end{equation}

If $\langle{u'{}^y u'{}^z}\rangle=0$ in the statistically steady state, then
\begin{equation}
	U^y=(\eta /\nuT)\, 2 \omega_{\rm{F}}^y H,
	\label{eq:meanUy}
\end{equation}
which corresponds to a mean flow in the direction of the rotation axis.


	We consider three values for the scale separation ratio, $\kf/k_1=5$, $15$, and $30$ and determine $\eta/\nu_{\rm{T}}$ using Eq.~(\ref{eq:meanUy}) by measuring $U^y$ and $H$ ($k_{\rm{f}}$: forcing wavenumber, $k_1$: wavenumber for system size).
We express time in terms of
\begin{equation}
	\tau=1/\urms\kf,
	\label{tau_def_num}
\end{equation}
which is also used as an estimate of the correlation time of the turbulence ($u_{\rm{rms}}$: root mean square velocity). Kinetic energy spectra, $E_{\rm K}(k,t)$, are normalized such that $\int E_{\rm K}(k,t)\,\dd k=\bra{\uu^2}/2$.

	All simulations are performed with the {\sc Pencil Code}\footnote{http://github.com/pencil-code/}, which uses a high-order finite difference method for solving the compressible hydrodynamic equations. We use a small Mach number so that the results are essentially the same as for a purely incompressible flow.

\subsection{Numerical results\label{sec:4B}}
The results are summarized in \Tab{tab:Summary}. All simulations show
that the sign of $\eta$ is {\it positive}. We find that $\eta/(\nuT\tau^2)$
is in the range of $O(10^{-2})$ to $O(10^{-1})$, depending on Reynolds
and Coriolis numbers (${\rm{Co}} = \omega_{\rm{F}} \tau$) as well as
scale separation. Run~A shows clear generation of a mean flow as seen
from Eq.~(\ref{eq:meanUy}). This equation is also used to determine
$\eta/(\nuT \tau^2)$ as the correlation coefficient in $U^y$ vs.\
$2\omega_{\rm{F}}^y H$; see the last column of \Tab{tab:Summary}.

\begin{table}[b!]
\caption{
Summary of DNS results. The Reynolds number is defined by $\Rey = u_{\rm{rms}} /(\nu k_{\rm{f}})$.
}\vspace{12pt}\centerline{\begin{tabular}{ccccccc}
Run & $\kf/k_1$ & $\Rey$  &  $\Co$  &  $\eta/(\nuT \tau^2)$ \\
\hline
A  &  15 &  60 &  0.74  &  0.22  \\ 
B1 &   5 & 150 &  2.6   &  0.27  \\ 
B2 &   5 & 460 &  1.7   &  0.27  \\ 
B3 &   5 & 980 &  1.6   &  0.51  \\ 
C1 &  30 & 18  &  0.63  &  0.50  \\ 
C2 &  30 & 80  &  0.55  &  0.03  \\ 
C3 &  30 &100  &  0.46  &  0.08  \\ 
\label{tab:Summary}\end{tabular}}
\end{table}

\subsubsection{Mean flows\label{sec:4B1}}
\begin{figure}[t!]\begin{center}
\includegraphics[width=.40\textwidth]{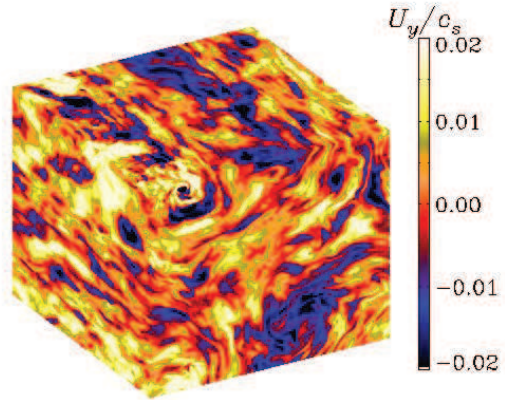}
\end{center}\caption[]{
Axial flow component $U^y$ on the periphery of the domain for Run~B2
with $\kf/k_1=5$ and $\Rey=460$.
}
\label{fig:velocity_contour}
\end{figure}

	As we see from Eq.~(\ref{eq:meanUy}), the large-scale flow is expected to be generated in the direction of the rotation vector $\mbox{\boldmath$\omega$}_{\rm{F}}$ (or the large-scale vorticity $\mbox{\boldmath$\Omega$}$) mediated by the helicity effect. The shape of the mean axial velocity component $U^y$ is shown in Fig.~\ref{fig:velocity_contour}. A clear flow pattern with positive and negative velocity is seen, which corresponds to the velocity distribution given by Eq.~(\ref{eq:meanUy}).

\begin{figure}[h!]\begin{center}
\includegraphics[width=.40\textwidth]{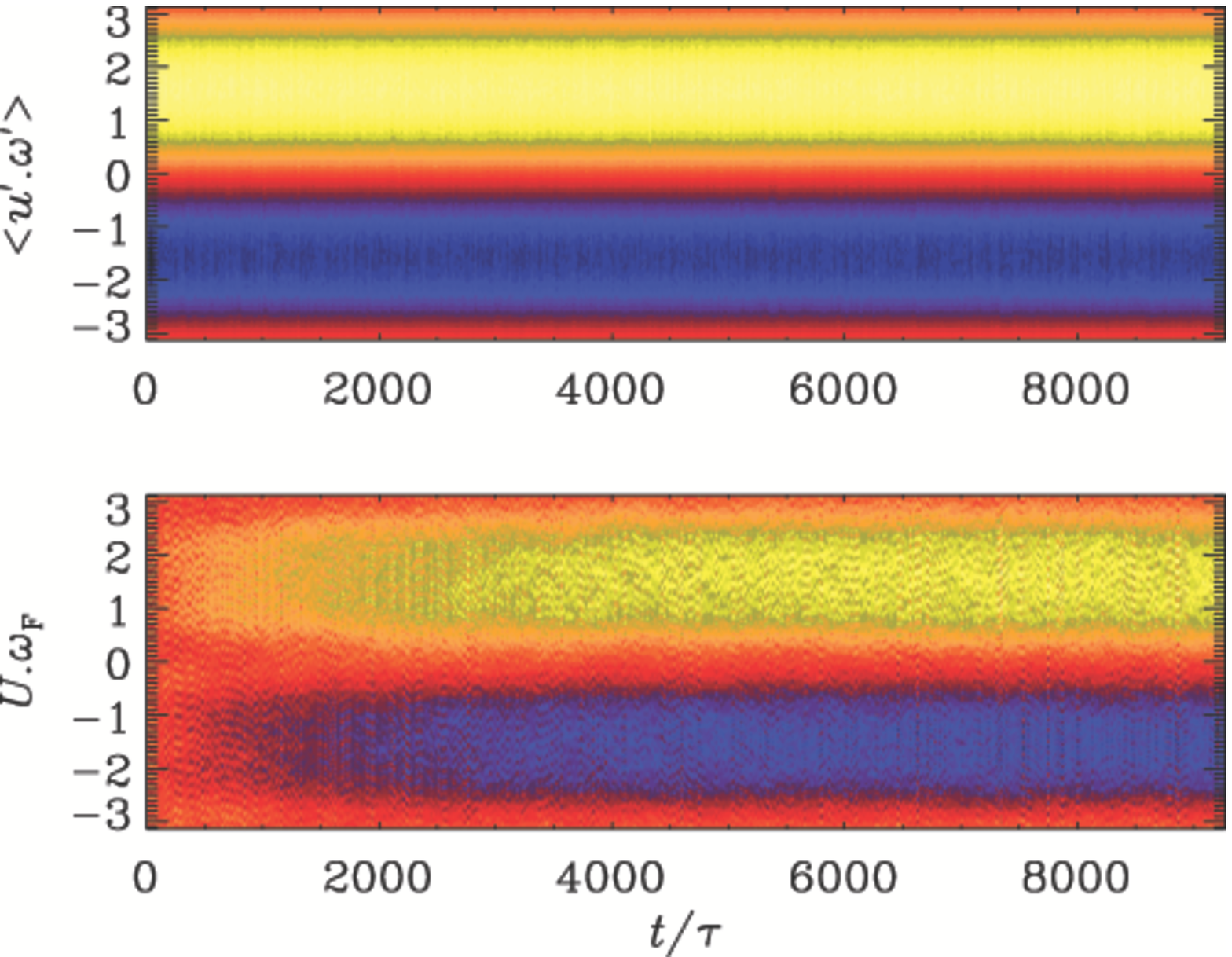}
\end{center}\caption[]{
Turbulent helicity $\langle {\bf{u}}' \cdot \mbox{\boldmath$\omega$}' \rangle$ (top) and mean-flow helicity ${\bf{U}} \cdot \mbox{\boldmath$\omega$}_{\rm{F}}$ (bottom) for Run~C1 with $\kf/k_1=30$ and $\Rey=18$.
}
\label{fig:turb_mean_helicity}
\end{figure}

	In Fig.~\ref{fig:turb_mean_helicity}, we show the temporal evolution of the turbulent helicity $\langle {\bf{u}}' \cdot \mbox{\boldmath$\omega$}' \rangle$ and the mean-flow helicity ${\bf{U}} \cdot \mbox{\boldmath$\omega$}_{\rm{F}}$. In this simulation, the turbulent helicity $\langle {\bf{u}}' \cdot \mbox{\boldmath$\omega$}' \rangle$ is sustained by the external forcing from the beginning of the simulation. Its spatial distribution reflects the forcing, which is proportional to $\sin k_1 z$ so that $H>0$ for $z>0$ and $H<0$ for $z<0$.  On the other hand, the mean-flow helicity ${\bf{U}} \cdot \mbox{\boldmath$\omega$}_{\rm{F}}$ is generated as the mean axial flow $U^y$ is induced by the inhomogeneous turbulent helicity effect. The magnitude of ${\bf{U}} \cdot \mbox{\boldmath$\omega$}_{\rm{F}}$ reaches an equilibrium value around $t / \tau = 2000$. Its spatial distribution is consistent with the direction of the induced axial flow $U^y$. 

\subsubsection{Reynolds stress tensor\label{sec:4B2}}
\begin{figure}[t!]
\includegraphics[width=.40\textwidth]{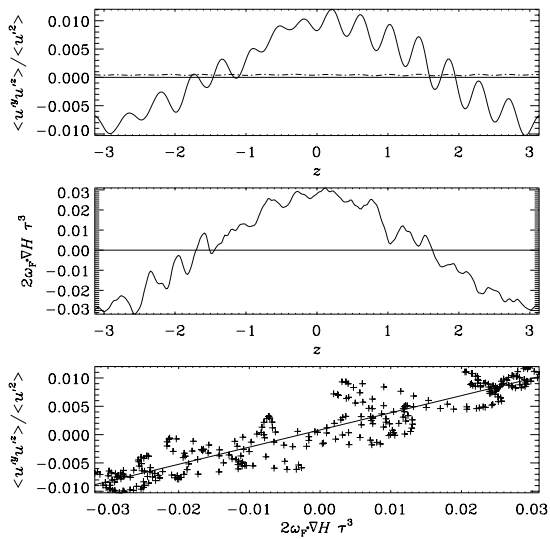}
\caption{Reynolds stress $\langle {u'{}^y u'{}^z} \rangle$ (top), helicity-effect term $(\nabla H)^z 2\omega_{\rm{F}}^y$ (middle), and their correlation (bottom) for Run~A with $\kf/k_1=15$ and $\Rey=60$ at the early stage of development (averaged over time from $t/\tau = 40$ to $200$). 
}
\label{fig:uzuy_gradH_corr}
\end{figure}

As noted in connection with Eqs.~(\ref{eq:Rey_strss_num_early}) and (\ref{eq:Rey_strss_num_dev}), at
the early stage of development, we have no large-scale flows. In
this case, the Reynolds stress should be represented only by the
$\mbox{\boldmath$\omega$}_{\rm{F}}$- or rotation-related terms in
Eqs.~(\ref{eq:Rey_strss_exprssn}) and (\ref{eq:Rey_strss_model}). First
we examine this early stage of development by taking an average over time
from $t/\tau = 40$ to $200$. The $y$-$z$ component of the Reynolds stress,
$\langle {u'{}^y u'{}^z} \rangle$ in the early stage is shown in the top
panel of Fig.~\ref{fig:uzuy_gradH_corr}. The spatially averaged magnitude
of the Reynolds stress is drawn with the dot dashed line. The top panel
shows that the peak magnitude of the Reynolds stress normalized by the
turbulent intensity $\langle u'{}^2 \rangle$ is about $0.01$. In the
middle panel of Fig.~\ref{fig:uzuy_gradH_corr}, the helicity-related
term $2 \omega_{\rm{F}}^y (\nabla H)^z$ is plotted against $z$; the
basic spatial profile reflects the counterpart of the turbulent helicity
schematically depicted in the center panel of Fig.~\ref{fig:setup}. The
spatial profile of the Reynolds stress $\langle {u'{}^y u'{}^z} \rangle$
is in remarkable agreement with the turbulent helicity gradient coupled
with the rotation, $2 \omega_{\rm{F}}^y (\nabla H)^z$. This agreement
confirms the validity of the model expression (\ref{eq:Rey_strss_model})
based on the theoretical result [Eq.~(\ref{eq:Rey_strss_exprssn})].

\begin{figure}[t!]
\includegraphics[width=.40\textwidth]{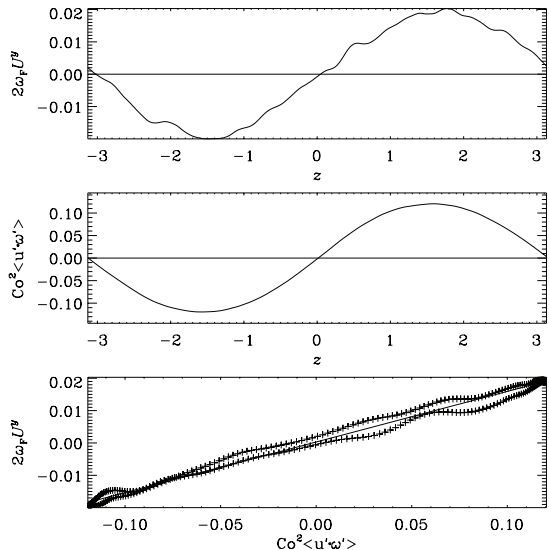}
\caption{Mean axial velocity $U^y$ (top), turbulent helicity multiplied by rotation $C_0^2 \langle {{\bf{u}}' \cdot \mbox{\boldmath$\omega$}'} \rangle = (2 \omega_{\rm{F}} \tau)^2 H$ (middle), and their correlation (bottom) for Run~A with $\kf/k_1=15$ and $\Rey=60$ at the developed equilibrium stage (averaged over time from $t/\tau =0$ to $2000$).}
\label{fig:uy_omega_yH_corr}
\end{figure}

	Next, we examine the correlation between the mean velocity and the helicity at the developed equilibrium stage reached around $t/\tau = 2000$. In the developed equilibrium state, the mean velocity should be related to the rotation and turbulent helicity as Eq.~(\ref{eq:meanUy}). In Fig.~{\ref{fig:uy_omega_yH_corr}}, we compare the mean axial velocity component $U^y$ and the turbulent helicity $H$. The correlation between the generated mean velocity and helicity is quite remarkable. This result also confirms the model expression for the Reynolds stress [Eq.~(\ref{eq:Rey_strss_model})].

	The ratio of the magnitudes of the helicity to the eddy-viscosity effects may be given by
\begin{eqnarray}
	\lefteqn{
	\frac{(\mbox{helicity effect})}{(\mbox{eddy-viscosity effect})} 
	= \frac{|\eta 2\omega_{\rm{F}} \nabla H|}{|\nu_{\rm{T}} \nabla U|}
	}\nonumber\\
	&& \hspace{30pt} \sim \frac{\eta}{\nu_{\rm{T}}} \frac{\Omega_\ast}{{\cal{S}}} {|\nabla H|}
	\sim \frac{\eta}{\nu_{\rm{T}} \tau^2}  \frac{\Omega_\ast}{{\cal{S}}},
	\label{eq:Rey_strss_hel_ratio}
\end{eqnarray}
where $\Omega_\ast$ is the magnitude of mean absolute vorticity $\mbox{\boldmath$\Omega$}_\ast (\equiv \mbox{\boldmath$\Omega$} + 2 \mbox{\boldmath$\omega$}_{\rm{F}}$) and ${\cal{S}}$ the magnitude of velocity strain. With the modeling of $\nu_{\rm{T}}$ [Eq.~(\ref{eq:nu_T_K_eps})] and $\eta$ [Eq.~(\ref{eq:eta_K_eps})], we see that $\eta/(\nu_{\rm{T}} \tau^2)$ in this ratio corresponds to the ratio of model constants $C_\eta / C_\nu$, which was estimated as
\begin{equation}
	\frac{\eta}{\nu_{\rm{T}} \tau^2} 
	\approx \frac{C_\eta}{C_\nu}
	= \frac{0.003}{0.09}
	\approx 0.03
	\label{eq:Ratio_value}
\end{equation}
in the turbulent swirling flow model simulation \cite{yok1993} (see Appendix~\ref{sec:appA2}). The present DNS results for $\eta / (\nu_{\rm{T}} \tau^2)$ listed in Table~\ref{tab:Summary} should be compared with this estimate Eq.~(\ref{eq:Ratio_value}). The value of $C_\eta / C_\nu$ utilized in the previous work \cite{yok1993} is in the range of the value of $\eta/(\nu_{\rm{T}} \tau^2)$ in the present DNSs. The agreement seems to be better in the case with a weaker rotation (smaller ${\rm{Co}}$) and a larger scale separation ($k_{\rm{f}} / k_1$). We should note that in the turbulent swirling flow, where the helicity turbulence model was applied, the ratio $\Omega_\ast / {\cal{S}}$ estimated from the scaled axial angular momentum flux was less than $0.2$ [$\int_0^a r U^\phi U^z 2\pi dr / (\pi a^2 U_{\rm{m}}^2) \simeq 0.18$ at inlet, $a$: pipe radius, $U^\phi$: circumferential velocity, $U^z$: axial velocity]. This is a much smaller rotation case as compared with the present DNSs. Also note that the Reynolds number of the swirling flow was much larger ($\sim 5 \times 10^{4}$) and that the turbulent helicity was provided by a large-scale swirling flow not by an external forcing.

\subsubsection{Spectra\label{sec:4B3}}
\begin{figure}[h!]\begin{center}
\includegraphics[width=.40\textwidth]{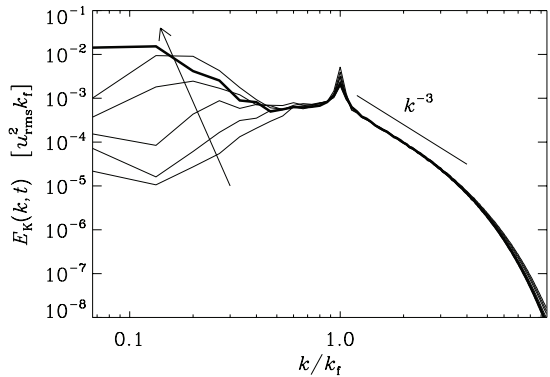}
\end{center}\caption[]{
	Inverse transfer seen in kinetic energy spectra at $t/\tau=100$,
200, 500, 1000, 2000, and 3500, for Run~A with $\kf/k_1=15$, $\Rey=60$,
$\Co=0.7$, and $288^3$ meshpoints.
The arrow denotes the temporal evolution.
The bold line indicates the last time in the plot.
}\label{pkt_per288_theta90a_kf15x}\end{figure}

	The spectra show a peak at the forcing wavenumber $\kf$ and, at early times, a second peak at $k/\kf\approx0.25$, which then gradually moves to smaller values of $k$; see \Fig{pkt_per288_theta90a_kf15x}.

\begin{figure}[h!]\begin{center}
\includegraphics[width=.40\textwidth]{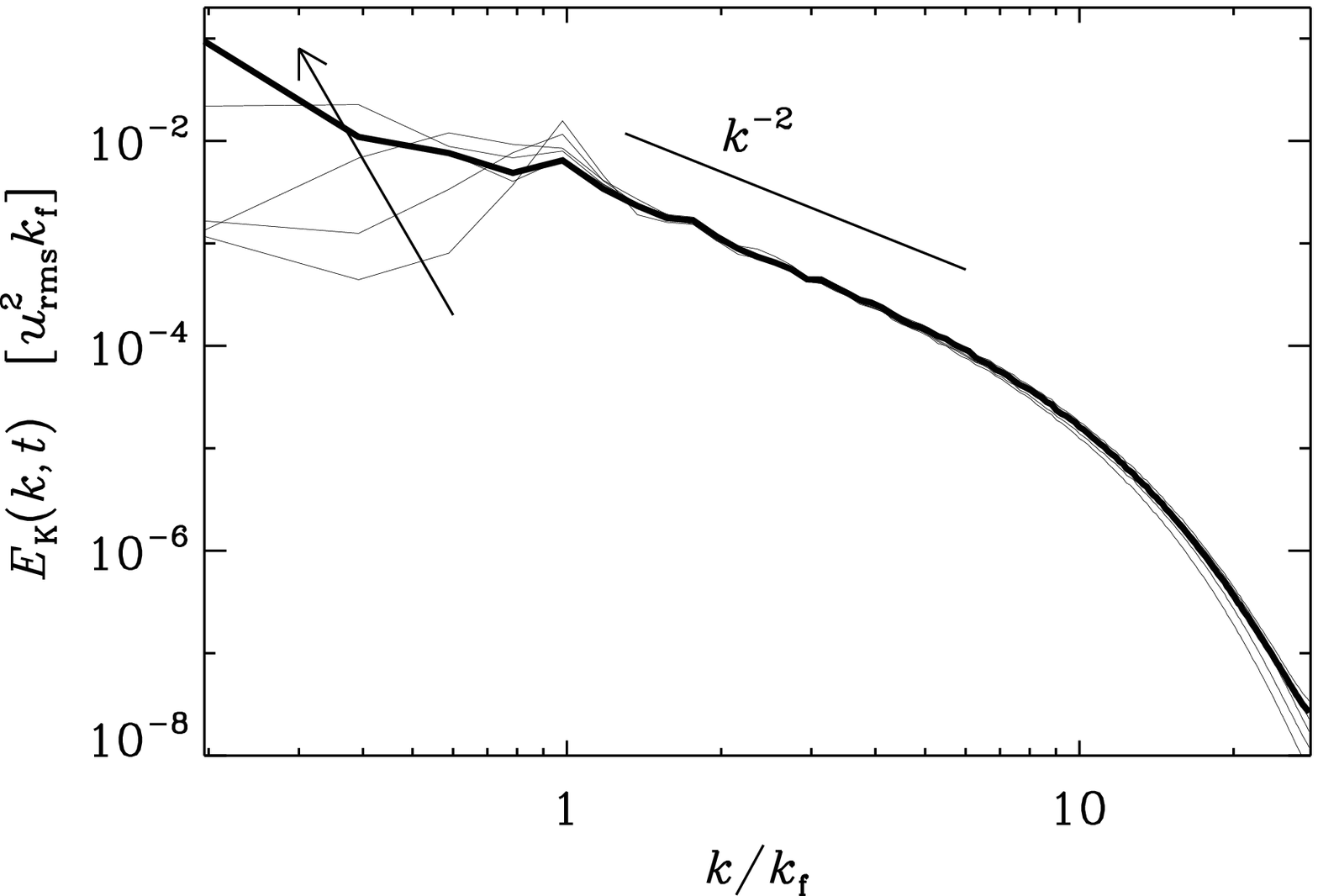}
\end{center}\caption[]{
Same as \Fig{pkt_per288_theta90a_kf15x}, but Run~B2 with $\kf/k_1=5$,
$\Rey=460$, and $t/\tau=50$, 100, 200, 500, and 1600, with
$\Co=0.7$, and $288^3$ meshpoints.
}\label{pkt_per288_thetam90a_kf5x_nu1em4}\end{figure}

	The inverse transfer behavior can also be seen at smaller scale separation. In \Fig{pkt_per288_thetam90a_kf5x_nu1em4} we show the spectra for Run~B2 with $\kf/k_1=5$ and $\Rey=460$. The flow takes the form of pairs of counterrotating vortices. This is shown in Fig.~\ref{fig:velocity_contour} for the same run.

\begin{figure}[h!]\begin{center}
\includegraphics[width=.40\textwidth]{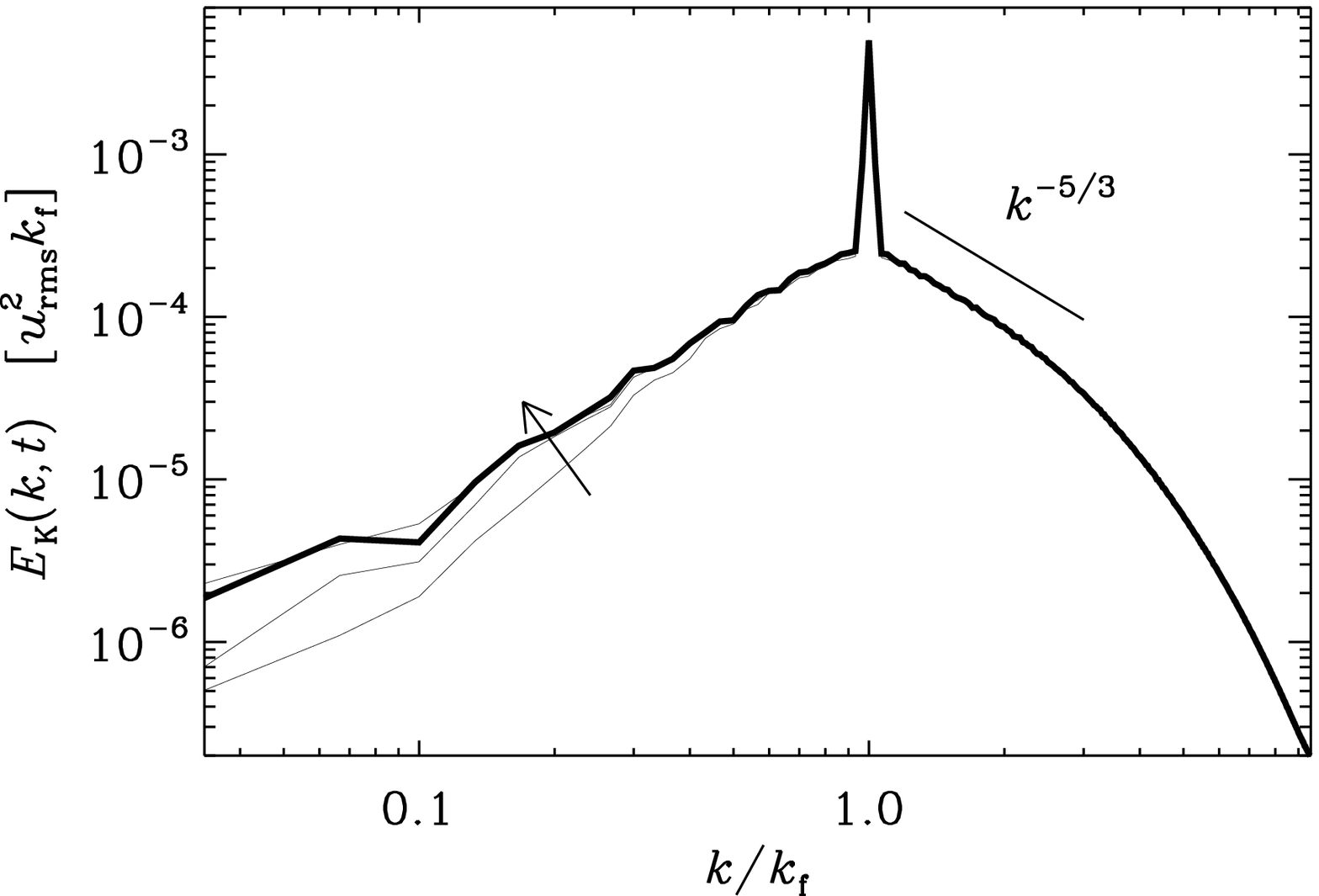}
\end{center}\caption[]{
Same as \Fig{pkt_per288_theta90a_kf15x}, but for Run~C2 with $\kf/k_1=30$,
$\Rey=80$, and $t/\tau=50$, 200, 1300, and 4700, with
$\Co=0.7$, and $576^3$ meshpoints.
}\label{pkt_per576_thetam90a_kf30x_nu5em5}\end{figure}

\begin{figure}[h!]\begin{center}
\includegraphics[width=.40\textwidth]{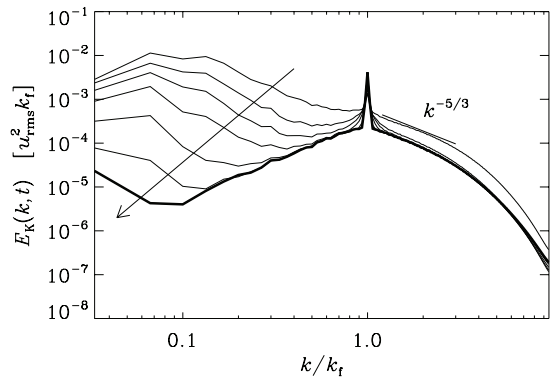}
\end{center}\caption[]{
Same as \Fig{pkt_per576_thetam90a_kf30x_nu5em5}, but for Run~C3 with a finite amplitude initial condition, giving
$\Rey=100$, at $t/\tau=20$, 100, 200, 500, 1000, and 1700.
Here, $\kf/k_1=30$, $\Co=0.5$, and $576^3$ meshpoints.
}\label{pkt_per576_thetam90a_kf30x_nu5em5_ini}
\end{figure}

	It is surprising at first sight that the inverse transfer behavior is {\it not} seen at larger scale separation. This is demonstrated in \Fig{pkt_per576_thetam90a_kf30x_nu5em5}, where we see the result for $\kf/k_1=30$ and $\Rey=80$. To test whether the inverse transfer might be the result of a subcritical bifurcation, we have performed an identical simulation, but with a different initial condition where an initial random flow with a $k^{-5/3}$ spectrum was used. The result is shown in \Fig{pkt_per576_thetam90a_kf30x_nu5em5_ini}, where we see that the initial power at large scales gradually disappears. This suggests that the large-scale flow only occurs at finite scale separation ratios of between 5 and 15.

These results show that the inverse cascade is less strong when the forcing is at smaller scales. This spectral behavior may be attributed to the finite size of the simulation box. Also this may be related to the fact that the relative importance of helicity to eddy-viscosity effects is scale dependent. The helicity effect (representing the transport suppression) is connected with the inverse cascade, whereas the eddy-viscosity effect (representing the transport enhancement) is connected with the normal cascade. From the spectral expression of the Reynolds stress [Eq.~(\ref{eq:Rey_strss_exprssn})], we estimate the relative importance of the helicity effect as
\begin{equation}
	\frac{(\mbox{helicity effect})}{(\mbox{eddy-viscosity effect})}
	\sim \frac{\Omega_\ast}{{\cal{S}}} \frac{|H(k,t)|}{kE(k,t)},
	\label{eq:relative_hel_effect}
\end{equation}
where $\Omega_\ast (= \sqrt{{\Omega_\ast}^{ab} \Omega_\ast^{ab}/2})$ and
${\cal{S}} (= \sqrt{{\cal{S}}^{ab} {\cal{S}}^{ab}/2})$ are the magnitudes
of the absolute vorticity and strain tensors, respectively. If turbulent
helicity obeys the same scaling as turbulent energy, $H(k,t) \sim E(k,t)
\propto k^{-p}$ ($p$: power index), we have $|H(k,t)|/[k E(k,t)] \propto
k^{-1}$ for the relative helicity. This suggests that the helicity
effect is less relevant as we go to smaller scales. In this sense, it
is important to see how the relative helicity $|H(k,t)|/[k E(k,t)]$
depends on the scale separation $k_{\rm{f}}/k_1$. Further examination
of these results is left for the interesting future work.

\section{Discussions\label{sec:5}}
Here we discuss the present mean-flow generation mechanism by inhomogeneous turbulent helicity in the context of vortex dynamo. Comparisons with some of the representative previous works on large-scale vorticity generation mechanism such as the Anisotropic Kinetic Alpha (AKA) effect \cite{fri1987} and the $\Lambda$ effect \cite{rue1980,rue1989} are presented in Appendix~\ref{sec:appB}.

\subsection{Vortex dynamo due to turbulent helicity\label{sec:5A}}
	By applying the curl operation to the mean momentum equation, we obtain the equation for the mean vorticity $\mbox{\boldmath$\Omega$} (= \nabla \times {\bf{U}})$ as Eq.~(\ref{eq:mean_vor_eq}). There the turbulent Vortex-Motive Force (VMF) ${\bf{V}}_{\rm{M}} = \langle {{\bf{u}}' \times \mbox{\boldmath$\omega$}'} \rangle$ represents the effect of turbulence in the $\mbox{\boldmath$\Omega$}$ equation. The turbulent VMF is directly related to the Reynolds stress $\mbox{\boldmath${\cal{R}}$} = \{ {{\cal{R}}^{ij}} \}$ [Eq.~(\ref{eq:rey_strss_def_2})] as
\begin{equation}
	V_{\rm{M}}^i
	= - \frac{\partial {\cal{R}}^{ij}}{\partial x^j}
	+ \frac{\partial K}{\partial x^i}.
	\label{eq:vmf_rey_strss_rel}
\end{equation}
Note that the second or $\nabla K$ term does not contribute to the vorticity generation at all since $\nabla \times \nabla K =0$. Substitution of $\mbox{\boldmath${\cal{R}}$}$ [Eq.~(\ref{eq:Rey_strss_exprssn})] into Eq.~(\ref{eq:vmf_rey_strss_rel}) gives the VMF expression as
\begin{equation}
	{\bf{V}}_{\rm{M}}
	= - D_\Gamma \left( {
		\mbox{\boldmath$\Omega$} + 2 \mbox{\boldmath$\omega$}_{\rm{F}}
	} \right)
	- \left[ {\left({
		\mbox{\boldmath$\Omega$} 
		+ 2 \mbox{\boldmath$\omega$}_{\rm{F}}
	} \right) \cdot \nabla} \right] \mbox{\boldmath$\Gamma$}
	- \nu_{\rm{T}} \nabla \times \mbox{\boldmath$\Omega$}
	\label{eq:vmf_exprssn}
\end{equation}
with
\begin{equation}
	D_\Gamma 
	= \nabla \cdot \mbox{\boldmath$\Gamma$}.
	\label{eq:D_gamma}
\end{equation}
The third or $\nu_{\rm{T}}$-related term in Eq.~(\ref{eq:vmf_exprssn}) is the turbulent diffusion of $\mbox{\boldmath$\Omega$}$, representing the destruction of the large-scale vorticity due to turbulence. The first and the second terms give a possibility of the mean vorticity generation due to the inhomogeneity of the turbulent helicity.

	We consider a situation where the angular velocity is much larger than the mean vorticity:
\begin{equation}
	|2 \mbox{\boldmath$\omega$}_{\rm{F}}|
	\gg |\mbox{\boldmath$\Omega$}|.
\end{equation}
In this case with the set-up we considered in the numerical simulation, the contribution from the second term vanishes since the direction of $\mbox{\boldmath$\omega$}_{\rm{F}}$ ($y$ direction) is perpendicular to the direction of the inhomogeneity ($z$ direction). We then have the turbulent VMF as
\begin{equation}
	{\bf{V}}_{\rm{M}}
	= - D_\Gamma 2\mbox{\boldmath$\omega$}_{\rm{F}}
	- \nu_{\rm{T}} \nabla \times \mbox{\boldmath$\Omega$}.
	\label{eq:vmf_simple_geometry}
\end{equation}

	In a special case with the spatial distribution of the turbulent kinetic helicity given by Eq.~(\ref{eq:H_profile}), we have
\begin{equation}
	D_\Gamma 
	\simeq C_\eta \tau \ell^2 \frac{\partial^2 H}{\partial z^2}
	= - 3 C_\eta \tau \ell^2 H_0 z.
	\label{eq:D_Gamma_special}
\end{equation}
In this case, the mean vorticity induction due to the first term in Eq.~(\ref{eq:vmf_simple_geometry}) is given by
\begin{subequations}\label{eq:vort_ind_simple}
\begin{equation}
	{\bf{I}}_{\rm{V}}
	= \nabla \times {\bf{V}}_{\rm{M}}
	= \nabla \times \left( {
		- D_\Gamma 2\mbox{\boldmath$\omega$}_{\rm{F}}
	} \right)
	= 2\mbox{\boldmath$\omega$}_{\rm{F}} \times \nabla D_\Gamma
	\label{eq:vort_ind_simple_vec}
\end{equation}
or in components,
\begin{equation}
	I_{\rm{V}}^\alpha
	= \epsilon^{\alpha ba} 
		2\omega_{\rm{F}}^b \frac{\partial D_\Gamma}{\partial x^a} .
	\label{eq:vort_ind_simple_comp}
\end{equation}
\end{subequations}
This leads to the mean vorticity generation in the $x$ direction as
\begin{equation}
	I_{\rm{V}}^x
	= 6 C_\eta \tau \ell^2 H_0 \omega_{\rm{F}}
	= 6 C_\eta \frac{K}{\varepsilon} \frac{K^3}{\varepsilon^2} 
		H_0 \omega_{\rm{F}}.
	\label{eq:vort_ind_y_simple}
\end{equation}
Note that the large-scale vorticity is generated in the direction perpendicular to both the directions of the angular velocity ($y$ direction) and of the turbulent helicity inhomogeneity ($z$ direction).

	Equation~(\ref{eq:vort_ind_y_simple}) shows that rotation coupled with inhomogeneous turbulent kinetic helicity generates a large-scale vorticity component that is not in the rotation direction, and the magnitude of generation in this particular case is uniform in space.
	
	Equation~(\ref{eq:vmf_simple_geometry}) should be compared with the equilibrium expression~(\ref{eq:meanUy}), which shows that the large-scale flow is generated in the direction of the rotation vector with a coefficient proportional to the turbulent helicity. The mean velocity equation is obtained by uncurling Eq.~(\ref{eq:mean_vor_eq}). We see from Eq.~(\ref{eq:vmf_simple_geometry}) that the mean velocity generation is in the direction of the rotation vector with the proportionality coefficient $-2 D_\Gamma$. In the special case we considered in the numerical simulation, $D_\Gamma$ may be expressed as Eq.~(\ref{eq:D_Gamma_special}), which corresponds to Eq.~(\ref{eq:meanUy}).

\subsection{Physical origin of the helicity effect\label{sec:5B}}
Since the Reynolds stress $\mbox{\boldmath${\cal{R}}$} = \{ {\cal{R}}^{ij}
\}$ is a rank two tensor, it is not so simple to draw an intuitive
physical picture of each component of the Reynolds stress. On the other
hand, the turbulent VMF ${\bf{V}}_{\rm{M}} = \langle {{\bf{u}}' \times
\mbox{\boldmath$\omega$}'} \rangle$ is a vector so that it is easier
to get a physical picture of ${\bf{V}}_{\rm{M}}$. The relationship
between ${\bf{V}}_{\rm{M}}$ and $\mbox{\boldmath${\cal{R}}$}$ is given
by Eq.~(\ref{eq:vmf_rey_strss_rel}). Here, we consider the turbulent
VMF in the mean vorticity equation, instead of the Reynolds stress in
the mean momentum equation, to understand the physical origin of the
present helicity effect.

A mean velocity induction constant in space does not contribute to the
generation of mean vorticity at all since $\mbox{\boldmath$\Omega$} =
\nabla \times {\bf{U}}$. So, we focus our attention on the inhomogeneous
helicity represented by $D_\Gamma = \nabla \cdot \mbox{\boldmath$\Gamma$}
\propto \nabla^2 H$ [Eq.~(\ref{eq:D_gamma})]. The Laplacian of $H$,
$\nabla^2 H$, quantifies how prominent the local $H$ is as compared with
the surrounding $H$ in average. The Laplacian may be estimated as
\begin{equation}
	\nabla^2 H \simeq - \frac{\delta H}{\ell^2} 
	= - \frac{\langle {{\bf{u}}' \cdot \delta \mbox{\boldmath$\omega$}'} \rangle}{\ell^2},
	\label{eq:laplace_H}
\end{equation}
where $\ell$ is the helicity variation scale, $\delta H$ is the helicity
variation relative to the average of the surroundings, and is $\delta
\mbox{\boldmath$\omega$}'$ the vorticity fluctuation associated with
$\delta H$. Positive $\delta H$ correspond to positive alignment
of $\delta\mbox{\boldmath$\omega$}'$ with the velocity fluctuation
${\bf{u}}'$ in the statistical sense ($\delta H = \langle {{\bf{u}}'
\cdot \delta \mbox{\boldmath$\omega$}'} \rangle > 0$) and vice versa.

\begin{figure}[t!]
\includegraphics[width=.40\textwidth]{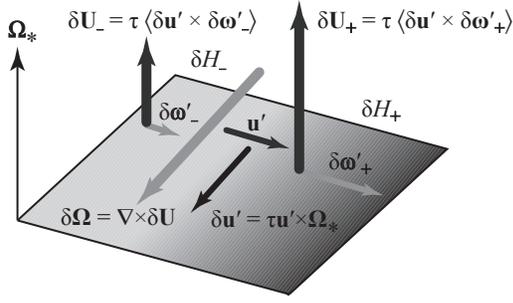}
\caption{Physical origin of the helicity effect.}
\label{fig:phys_hel_effect} 
\end{figure}

We consider a fluid element fluctuating with ${\bf{u}}'$ in
the mean absolute vorticity $\mbox{\boldmath$\Omega$}_\ast$
(Fig.~\ref{fig:phys_hel_effect}). We further assume an
inhomogeneous helicity density with $\nabla^2 H < 0$ [i.e., $\delta H > 0$
according to Eq.~(\ref{eq:laplace_H})] and the relative helicity
variation $\delta H$ varies in space ($\delta H_+ > \delta H_-$ in
Fig.~\ref{fig:phys_hel_effect}). Equation~(\ref{eq:laplace_H}) indicates
that $\delta \mbox{\boldmath$\omega$}'$ is statistically parallel to
${\bf{u}}'$ ($\langle {{\bf{u}}' \cdot \delta \mbox{\boldmath$\omega$}'}
\rangle > 0$) although each realization is more random. In this figure,
for the sake of simplicity, the direction of ${\bf{u}}'$ is drawn in
the direction parallel to the gradient of $\delta H$. Note that the
present argument applies for any ${\bf{u}}'$ direction with respect to the
gradient of $\delta H$. For a given ${\bf{u}}'$, the magnitude of $\delta
\mbox{\boldmath$\omega$}'$ reflects that of $\delta H$.
It is also worthwhile to remark that the spatial variation of $\nabla^2 H$ produces a
non-uniform flow necessary for the induction of large-scale vorticity.

In the presence of absolute vorticity $\mbox{\boldmath$\Omega$}_\ast$,
a fluid element moving with ${\bf{u}}'$ is subject to the
Coriolis-like force to induce a flow modulation $\delta {\bf{u}}'
= \tau {\bf{u}}' \times \mbox{\boldmath$\Omega$}_\ast$. Then, we
have a contribution to the VMF and consequently to the mean velocity
induction $\delta {\bf{U}} = \tau \langle {\delta{\bf{u}}' \times \delta
\mbox{\boldmath$\omega$}'} \rangle$, whose direction is parallel to the
mean absolute vorticity when $\nabla^2 H < 0$. As a result,
mean vorticity is generated as $\delta \mbox{\boldmath$\Omega$}
= \nabla \times \delta{\bf{U}}$, which is in the direction of
$\mbox{\boldmath$\Omega$}_\ast \times \nabla (\nabla^2 H)$. This is in
agreement with Eq.~(\ref{eq:vort_ind_simple}). Note that the direction
of the gradient of $\nabla^2 H$ is opposite to that of $\delta H$ as
in Eq.~(\ref{eq:laplace_H}).

These arguments show that the basic elements of the present helicity
effect are (i) local angular momentum conservation represented by
the Coriolis force; and (ii) the presence of an inhomogeneous turbulent
helicity.

\section{Conclusions\label{sec:6}}
	The effect of kinetic helicity in the turbulent momentum transport was investigated. We assumed the generic statistical properties for the basic or lowest-order fields of homogeneous isotropic and non-mirrorsymmetric turbulence. It was shown that, as a higher-order or inhomogeneity contribution, the turbulent helicity gradient naturally enters the Reynolds-stress expression as the coupling coefficient of the mean vorticity and/or angular velocity. The inhomogeneous turbulent helicity coupled with the mean vorticity or rotation may contribute to the generation of large-scale flow. This mechanism was examined with the aid of DNSs of rotating turbulence with non-uniform helicity sustained by an external forcing. The numerical result showed a good correlation between the Reynolds stress and the helicity inhomogeneity coupled with the rotation in this simple flow geometry. This confirmed the inhomogeneous helicity effect in large-scale flow generation; a large-scale vortical motion is generated by inhomogeneous turbulent helicity in the presence of rotation. Unlike other vorticity-generation mechanisms such as baroclinicity, the present helicity effect can work even in incompressible turbulence. Since non-uniform turbulent helicity is easily generated by rotation in the presence of boundaries, this large-scale flow generation mechanism is expected to be ubiquitous in astro- and geo-physical phenomena. At the same time, density stratification, as well as rotation, is one of the main factors that produce turbulent helicity. In this sense, the large-scale flow generation due to the inhomogeneous helicity effect in compressible turbulence will provide an interesting subject for future investigation.

\begin{acknowledgments}
	The authors would like to thank Jim Wallace and Fazle Hussain for invaluable comments on the experimental and numerical studies of helicity effects in turbulent flows. Their thanks are also due to Simon Candelaresi for useful discussions on vortex dynamo. They are grateful to Robert Rubinstein and unanimous referees for suggestions substantially improving the presentation of manuscript. Support by the NORDITA Program on Magnetic Reconnection in Plasmas (2015) is also acknowledged. This work was supported by the Japan Society for the Promotion of Science (JSPS) Grants-in-Aid for Scientific Research (No.\ 24540228).
\end{acknowledgments}

\appendix

\section{Helicity turbulence model and its application to a swirling flow}\label{sec:appA}
	In the present model, the turbulent helicity in a gradient form shows up in the Reynolds-stress expression. As this consequence, the transport equation of the turbulent helicity should be also considered. This is a natural extension of the usual $K - \varepsilon$ model, and is called the helicity turbulence model. This model was already applied to a turbulent swirling pipe flow in \cite{yok1993}. Here, we outline the basic model structure and present some of the  application results that are relevant to the present paper. For the details of the modeling and its results in a turbulent swirling flow, see \cite{yok1993}.

\subsection{Helicity turbulence model\label{sec:appA1}}
	In addition to the mean momentum equation (\ref{eq:mean_vel_eq}) with the Reynolds stress expression (\ref{eq:Rey_strss_model}), the evolution equations of the turbulent statistical quantities, $K$, $\varepsilon$, and $H$, should be simultaneously considered. From the fluctuation velocity and vorticity equations, the equations of $K$ and $H$ are written in the same form as
\begin{equation}
	\left( {
		\frac{\partial}{\partial t}
		+ {\bf{U}} \cdot \nabla
	} \right) G
	= P_G
	- \varepsilon_G
	+ \nabla \cdot {\bf{T}}_G,
	\label{eq:G_eq}
\end{equation}
where $P_G$, $\varepsilon_G$, and ${\bf{T}}_G$ are the production, dissipation, and the transport rates of the quantity $G = (K, H)$. They are defined and modeled by
\begin{subequations}\label{eq:K_eq}
\begin{equation}
	P_K = - \left\langle {u'{}^a u'{}^b} \right\rangle \frac{\partial U^b}{\partial x^a},
	\label{eq:P_K_eq}
\end{equation}
\begin{equation}
	\varepsilon_K = \varepsilon,
	\label{eq:eps_K_eq}
\end{equation}
\begin{equation}
	{\bf{T}}_K
	= - \left\langle { \left( {\frac{{\bf{u}}'{}^2}{2} + p'}\right) {\bf{u}}'
	+ \nu \nabla K
	} \right\rangle 
	= \frac{\nu_{\rm{T}}}{\sigma_K} \nabla K.
	\label{eq:T_K_eq}
\end{equation}
\end{subequations}
\begin{subequations}\label{eq:H_eq}
\begin{equation}
	P_H = - \left\langle {u'{}^a u'{}^b} \right\rangle
		\frac{\partial \Omega^b}{\partial x^a}
		- \left( {\Omega^a + 2 \omega_{\rm{F}}^a} \right)
		\frac{\partial}{\partial x^b}\left\langle {u'{}^a u'{}^b} \right\rangle,
	\label{eq:P_H_eq}
\end{equation}
\begin{equation}
	\varepsilon_H = 2\nu \left\langle {
		\frac{\partial u'{}^a}{\partial x^b}
		\frac{\partial \omega'{}^a}{\partial x^b}
	} \right\rangle
	= C_H \frac{\varepsilon}{K} H,
	\label{eq:eps_H_eq}
\end{equation}
\begin{eqnarray}
	{\bf{T}}_H
	&=& K \left( {
		\mbox{\boldmath$\Omega$} + 2 \mbox{\boldmath$\omega$}_{\rm{F}}
	} \right)
	- \left\langle {
		({\bf{u}}' \cdot \mbox{\boldmath$\omega$}') {\bf{u}}'} \right\rangle
	+ \left\langle {
	\left( {\frac{{\bf{u}}'{}^2}{2} - p'} \right) \mbox{\boldmath$\omega$}'
	} \right\rangle
	\nonumber\\ 
	&=& K \left( {
		\mbox{\boldmath$\Omega$} + 2 \mbox{\boldmath$\omega$}_{\rm{F}}
	} \right)
	+ \frac{\nu_{\rm{T}}}{\sigma_H} \nabla H.
	\label{eq:T_H_eq}
\end{eqnarray}
\end{subequations}
The model equation of the energy dissipation rate $\varepsilon$ is given by
\begin{equation}
	\left( {
	\frac{\partial}{\partial t} + {\bf{U}} \cdot \nabla
	} \right) \varepsilon
	= C_{\varepsilon 1} \frac{\varepsilon}{K} P_K
	- C_{\varepsilon 2} \frac{\varepsilon}{K} \varepsilon
	+ \nabla \cdot \left( {
		\frac{\nu_{\rm{T}}}{\sigma_\varepsilon} \nabla \varepsilon
	} \right).
	\label{eq:eps_eq}
\end{equation}
Here $\sigma_K$, $C_H$, $\sigma_H$, $C_{\varepsilon 1}$, $C_{\varepsilon 2}$, and $\sigma_\varepsilon$ are model constants. A system of equations: the mean velocity equation with the Reynolds stress [Eq.~(\ref{eq:Rey_strss_model})] with the transport coefficients $\nu_{\rm{T}}$ [Eq.~(\ref{eq:nu_T_K_eps})] and $\eta$ [Eq.~(\ref{eq:eta_K_eps})] and the equations of the turbulence statistical quantities, Eqs.~(\ref{eq:G_eq}) and (\ref{eq:eps_eq}), constitutes a turbulence model. This is a closed system of equations since the turbulent transport coefficients $\nu_{\rm{T}}$ and $\eta$ or $\mbox{\boldmath$\Gamma$}$ are expressed in terms of the turbulent statistical quantities, whose evolutions are solved simultaneously with the mean-field equation. This model is reduced to the usual $K -\varepsilon$ model in the absence of the turbulent helicity ($H=0$). This is a natural extension of the $K-\varepsilon$ model, and we may call it the helicity turbulence model.

	We can also construct the transport equation for the dissipation rate of the turbulent helicity, $\varepsilon_H$, from the fundamental equations \cite{yok2016a}. However, for the sake of simplicity, we adopt a simplest possible algebraic model for $\varepsilon_H$ as Eq.~(\ref{eq:eps_H_eq}).

\subsection{Application to turbulent swirling flow\label{sec:appA2}}
	The helicity turbulence model was applied to a turbulent swirling pipe flow. We consider a swirling pipe flow in cylindrical coordinates $(r, \theta, z)$ with the central pipe axis being in the $z$ direction. In a turbulent pipe flow, the mean axial velocity $U^z$ shows a very flat radial profile, which is one of the consequences of the enhanced momentum transport due to turbulence. No mean-flow inhomogeneities can be allowed in a highly turbulent state in the main portion of the flow. However this situation drastically changes if we apply a swirling or circumferential flow $U^\theta$ in addition to the axial flow. The flat profile of the mean axial velocity changes to inhomogeneous one with the dent or deceleration in the central axis region. It has been well known that the usual $K-\varepsilon$ model applied to the turbulent swirling flow completely fails in reproducing this dent profile. Due to the strong momentum transport arising from turbulent viscosity, the dent profile imposed in the inlet region rapidly decays and the usual flat profile is immediately realized in the numerical simulation \cite{kob1987}. On the other hand, it is ubiquitously observed in the experiments that the swirling velocity and the dent profile of the mean axial velocity in the central axis region is sustained further downstream \cite{kit1991,ste1995}. This is one of the main drawbacks of the $K-\varepsilon$-type turbulence model applied to swirling pipe flow.

\begin{figure}[t!]
\includegraphics[width=.40\textwidth]{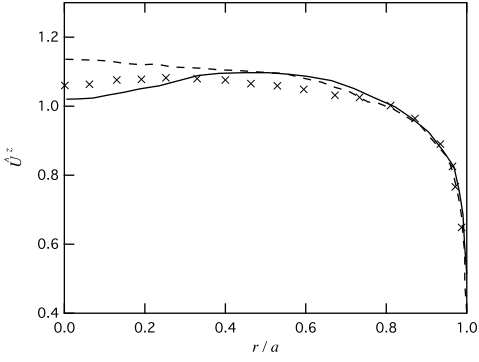}
\caption{Mean axial velocity profile in the turbulent swirling flow. The normalized mean axial velocity $\hat{U}^z$ ($= U^z / U_{\rm{m}}$, $U_{\rm{m}}$: bulk velocity defined by $U_{\rm{m}} = \int_0^a U^z 2\pi r dr / \pi a^2$) is plotted against the scaled radius $r/a$ ($a$: pipe radius): $\times$, experiment; -----, $K-\epsilon-H$ model; -- --, standard $K-\epsilon$ model. Redrawn from \citep{yok1993}.}
\label{fig:axial_vel_swirl} 
\end{figure}

	The presence of the rotational or circumferential motion gives a breakage of mirror symmetry between the parallel and antiparallel directions along the central axis. This suggests that it makes sense to consider the turbulent helicity in a swirling flow configuration. With this thought, the helicity turbulence model mentioned above was applied numerically to a swirling pipe flow \cite{yok1993}. Due to the helicity effect coupled with the mean vorticity represented by the second term in Eq.~(\ref{eq:Rey_strss_model}), the enhanced momentum transport due to turbulence is effectively suppressed. As shown in Fig.~\ref{fig:axial_vel_swirl}, the inhomogeneous dent profile of the mean axial velocity in the central axis region imposed as the inlet boundary condition is sustained up to downstream. This is in marked contrast with the result obtained with the usual $K-\varepsilon$ model, where the dent profile immediately disappears due to a strong effect of the turbulent viscosity represented by the first term in Eq.~(\ref{eq:Rey_strss_model}). 
	
If we adopt a slightly smaller model constant for the helicity-related
term in Eq.~(\ref{eq:Rey_strss_model}), say, $C_\eta = 0.002$, the
dent at the central axis decreases to be closer to the experimental
profile. However, we did not need to adopt such an approach in \cite{yok1993}
since our purpose was not to pursue the model achievement by optimizing
the model constants, but to show the basic tendency of the model; the
suppression of turbulent momentum transport by the helicity effect.

	This numerical simulation with the aid of turbulence or closure model confirms that the inhomogeneity of the turbulent helicity plays a key role in the transport suppression in turbulence. The enhanced momentum transport due to the turbulent or eddy viscosity can be effectively suppressed by the inhomogeneous turbulent helicity effect coupled with the mean vortical motion.

\section{Comparison with related effects\label{sec:appB}}
	Here we list two representative models previously studied as a vortex generation mechanism: (i) the Anisotropic Kinetic Alpha (AKA) effect \cite{fri1987}; and (ii) the $\Lambda$ effect \cite{rue1980,rue1989}; and compare them with the present helicity one.

\subsection{AKA effect\label{sec:appC1}}
	The Anisotropic Kinetic Alpha (AKA) effect is a three-dimensional large-scale instability of the hydrodynamic flow lacking parity invariance. The lack of parity invariance is broader concept than the helicity. The basic procedure deriving the AKA effect is as follows. (i) Solve the equation for the perturbed small-scale field:
\begin{equation}
	\frac{\partial u'{}^\alpha}{\partial t}
	+ U^a \frac{\partial u'{}^\alpha}{\partial x^a}
	= - \frac{\partial}{\partial x^a}(u'{}^a u'{}^\alpha)
	- \frac{\partial p'}{\partial x^\alpha}
	+ \nu \frac{\partial^2 u'{}^\alpha}{\partial x^a \partial x^a}
	+ f^\alpha
	\label{eq:aka_fluct_eq}
\end{equation}
[${\bf{f}}({\bf{x}},t)$: time-dependent space-and-time-periodic external force] with the assumption that the turbulent or small-scale Reynolds number is small. Here the mean velocity is assumed to be uniform over the small-scale length and time scales. (ii) Calculate the Reynolds stress ${\cal{R}}^{\alpha\beta} = \langle {u'{}^\alpha u'{}^\beta} \rangle$. If the mean velocity is small, the Reynolds stress can be expressed by using the Taylor expansion as
\begin{equation}
	{\cal{R}}^{\alpha\beta}
	= \langle {u'_0{}^\alpha u'_0{}^\beta} \rangle
	+ U^\ell \left. {
		\frac{\partial \langle{u'{}^\alpha u'{}^\beta}\rangle}{\partial U^\ell}
	} \right|_{U=0}
	+ \cdots,
	\label{eq:aka_rey_strss_tylr_exp}
\end{equation}
where ${\bf{u}}'_0$ is the basic turbulence field driven by ${\bf{f}}({\bf{x}},t)$, which satisfies an incompressible Navier--Stokes equation as
\begin{equation}
	\frac{\partial u'_{0}{}^\alpha}{\partial t}
	+  \frac{\partial}{\partial x^a}(u'_{0}{}^a u'_{0}{}^\alpha)
	= - \frac{\partial p'_{0}}{\partial x^\alpha}
	+ \nu \frac{\partial^2 u'_{0}{}^\alpha}{\partial x^a \partial x^a}
	+ f^\alpha.
	\label{eq:aka_basic_fld_eq}
\end{equation}
(iii) Substitute the Reynolds stress expression into the mean-velocity equation:
\begin{equation}
	\frac{\partial U^\alpha}{\partial t}
	+ U^a \frac{\partial U^\alpha}{\partial x^a}
	= - \frac{\partial R^{a \alpha}}{\partial x^a}
	- \frac{\partial P}{\partial x^\alpha}
	+ \nu \frac{\partial^2 U^\alpha}{\partial x^a \partial x^a}.
	\label{eq:aka_mean_fld_eq}
\end{equation}
If we adopt expansion (\ref{eq:aka_rey_strss_tylr_exp}), the equation for the large-scale motion becomes
\begin{equation}
	\frac{\partial U^\alpha}{\partial t}
	= \alpha^{\alpha a \ell} \frac{\partial U^\ell}{\partial x^a}
	- \frac{\partial P}{\partial x^\alpha}
	+ \nu \frac{\partial^2 U^\alpha}{\partial x^a \partial x^a},
	\label{eq:aka_lin_eq}
\end{equation}
where
\begin{equation}
	\alpha^{\alpha a \ell} 
	= - \left. { \frac{\partial \langle {
		u'{}^\alpha u'{}^a
	} \rangle}{\partial U^\ell}
	} \right|_{U=0}.
	\label{eq:aka_alpha_def}
\end{equation}
Equation~(\ref{eq:aka_lin_eq}) is referred to as the linear AKA equation. In the nonlinear regime, the AKA equation is written as
\begin{equation}
	\frac{\partial U^\alpha}{\partial t}
	+ \frac{\partial}{\partial x^a} U^a U^\alpha
	= \alpha^{\alpha a \ell} \frac{\partial U^\ell}{\partial x^a}
	- \frac{\partial P}{\partial x^\alpha}
	+ \nu \frac{\partial^2 U^\alpha}{\partial x^a \partial x^a}.
	\label{eq:aka_non_eq}
\end{equation}

	Note that, in Eq.~(\ref{eq:aka_fluct_eq}), the mean velocity-gradient term  $-u'{}^a (\partial U^\alpha / \partial x^a)$ is neglected. Also the Reynolds-stress term $(\partial / \partial x^a) {\cal{R}}^{a \alpha}$ is dropped. Such terms may be neglected in low-Reynolds-number flow, but play an important role in the evolution of fluctuations in realistic fully-developed turbulence with mean-field inhomogeneities.
	
	As has been mentioned, the AKA effect operates basically only at low Reynolds numbers \cite{BvR01}, when the eddy or turbulent viscosity is negligible in the mean momentum equation. This is reflected in Eqs.~(\ref{eq:aka_lin_eq}) and (\ref{eq:aka_non_eq}), where the molecular viscosity $\nu$ is the only mechanism that provides diffusive transport of the mean momentum. This point is in marked contrast both to the present helicity effect and to the $\Lambda$ effect.

\subsection{$\Lambda$ effect\label{sec:appC2}}
	The $\Lambda$ effect is a turbulence contribution to the Reynolds stress arising from anisotropy in the turbulence. A similar effect arises in magnetohydrodynamic (MHD) turbulence, where the turbulent Electro-Motive Force (EMF) in the mean magnetic induction equation, $\langle {{\bf{u}}' \times {\bf{b}}'} \rangle$, is expressed as a linear combination of the functionals of the mean magnetic field as \cite{KR80}
\begin{equation}
	\left\langle {{\bf{u}}' \times {\bf{b}}'} \right\rangle^\alpha
	= \cdots + \alpha^{\alpha a} B^a 
	+ \cdots.
	\label{eq:emf_fnctnl}
\end{equation}
By analogy with the EMF in the MHD turbulence, the Reynolds stress in the mean momentum equation in the hydrodynamic turbulence, $\langle {u'{}^\alpha u'{}^\beta} \rangle$, is supposed to be expressed in terms of a linear combination of the functionals of the angular velocity $\mbox{\boldmath$\omega$}_{\rm{F}}$ as
\begin{equation}
	\langle {u'{}^\alpha u'{}^\beta} \rangle
	= \cdots + \Lambda^{\alpha\beta a} \omega_{\rm{F}}^a
	+ \cdots.
	\label{eq:rey_strss_fnctnl}
\end{equation}
The transport coefficient $\Lambda^{\alpha\beta\gamma}$ is symmetric with respect to its indices $\alpha$ and $\beta$, and should have positive parity since both the Reynolds stress $\langle {u'{}^\alpha u'{}^\beta} \rangle$ and the angular velocity $\mbox{\boldmath$\omega$}_{\rm{F}}$ have positive parity as in Eqs.~(\ref{eq:Rey_strss_parity}) and (\ref{eq:Omega_parity}).

	The Reynolds-stress expression usually contains the eddy-viscosity or diffusive part as
\begin{equation}
	\langle {u'{}^\alpha u'{}^\beta} \rangle
	= \cdots - \nu_{\rm{T}} {\cal{S}}^{\alpha\beta} + \cdots,
	\label{eq:eddy_visc_model}
\end{equation}
which always enhances the momentum transport and smoothes out the inhomogeneous flow structures. On the other hand, the $\Lambda$ effect may generate a non-uniform flow.

	The $\Lambda$ effect is similar to the present helicity effect in that both effects occur in the strong turbulence regime, where the primary turbulence effect is the transport enhancement represented by the eddy or turbulent viscosity. This is in strong contrast with the AKA effect which works only in low Reynolds number turbulence, when we do not have any turbulent diffusive effects. 
	
	On the other hand, in the formulation of the $\Lambda$ effect, the form of the Reynolds stress, Eq.~(\ref{eq:rey_strss_fnctnl}), is assumed through the symmetry arguments on the Reynolds stress [Eq.~(\ref{eq:Rey_strss_parity})] and the angular velocity [Eq.~(\ref{eq:Omega_parity})]. Also in the $\Lambda$-effect formulation, the expression for the coefficient tensor $\Lambda^{\alpha\beta\gamma}$ is constructed on the basis of the considerations of parity symmetry (or asymmetry), anisotropy, boundary conditions, etc. As for the preferred direction leading to anisotropy, the radial or ``gravity'' direction $\hat{\bf{g}}$ ($= {\bf{g}} / |{\bf{g}}|$, ${\bf{g}}$: the gravitational acceleration vector) is usually adopted in the context of stellar differential rotation. As a result, the ``rigid'' rotation-related part of $\Lambda^{\alpha\beta\gamma}$ is written as \cite{rue1989}
\begin{equation}
	\Lambda^{\alpha\beta\gamma}
	= \Lambda_{\rm{V}} \left( {
		\epsilon^{\alpha a \gamma} {\hat{g}}^\beta
		+ \epsilon^{\beta a \gamma} {\hat{g}}^\alpha
	} \right) {\hat{g}}^a
	\label{eq:aka_Lambda_rigid}
\end{equation}
($\Lambda_{\rm{V}}$: magnitude of the vertical $\Lambda$ effect).

	By contrast, in the present helicity-effect formulation, the dependence of the Reynolds stress is derived through the mean-field inhomogeneities from the fundamental equations without resorting to any prescribed form of the Reynolds stress. Note that it is only the statistical property of the basic fields expressed in Eqs.~(\ref{eq:iso_nonmirror}) and (\ref{eq:G_iso}) that is assumed in the present formulation, and the dependence of the Reynolds stress on the mean-field inhomogeneities automatically appear through the differential expansion of the field quantities [Eqs.~(\ref{eq:diff_exp}) and (\ref{eq:two-scale_diff_exp})].

	These features of the present formulation constitute considerable advantages in treating cases with several kinds of mean-field inhomogeneities. The cases of treating the Reynolds stress and the turbulent EMF in inhomogeneous turbulence are such examples. In the presence of a non-uniform mean flow, the turbulent EMF should depend on the mean vorticity (antisymmetric part of the mean velocity shear) in addition to the usual dependence on the mean magnetic field \cite{yok2013,yok2016b}. Similarly, we should consider the mean magnetic-field shear in addition to the eddy-viscosity (mean velocity strain) and the helicity or $\Lambda$ (vorticity and/or angular velocity) terms \cite{yos1990,yok2013}.



\begin{thebibliography}{99}

\bibitem{par1955}
E. N. Parker,
``Hydromagnetic dynamo models,''
Astrophys. J. {\bf{122}}, 293-314 (1955).

\bibitem{bra2005}
A. Brandenburg and K. Subramanian,
``Astrophysical magnetic fields and nonlinear dynamo theory,''
Phys.\ Reports {\bf{417}}, 1-209 (2005).

\bibitem{yok2013}
N. Yokoi,
``Cross helicity and related dynamo,''
Geophys.\ Astrophys.\ Fluid Mech.\ {\bf{107}}, 114-184 (2013).

\bibitem{wed2012}
S. Wedemeyer-B\"{o}hm, E. Scullion, O. Steiner, L. Rouppe van der Voort, J. de la Cruz Rodriguez, V. Fedun, and R. Erd\'{e}lyi,
``Magnetic tornadoes as energy channels into the solar corona,''
Nature {\bf{486}}, 504-508 (2012).

\bibitem{mof1992}
H. K. Moffatt,
``Relaxation under topological constraints,''
in {\it Topological Aspects of Dynamics of Fluids and Plasmas}
edited by H. K. Moffatt, G. M. Zaslavsky, P. Comte, and M. Tabor (Kluwer Academic Publishers, Dordrecht, 1992), pp.~3-28.

\bibitem{yok1993}
N. Yokoi and A. Yoshizawa,
``Statistical analysis of the effects of helicity in inhomogeneous turbulence,''
Phys.\ Fluids A {\bf{5}},  464-477 (1993).

\bibitem{bri1973}
A. Brissaud, U. Frisch, J. L\'{e}orat, M. Lesieur, and A. Mazure,
``Helicity cascade in fully developed isotropic turbulence,''
Phys.\ Fluids {\bf{16}}, 1366-1367 (1973).

\bibitem{kra1973}
R. H. Kraichnan,
``Helical turbulence and absolute equilibrium,''
J. Fluid Mech.\ {\bf{59}}, 745-752 (1973).

\bibitem{and1977}
J. C. Andr\'{e} and M. Lesieur,
``Influence of helicity on the evolution of isotropic turbulence at high Reynolds number,''
J. Fluid Mech.\ {\bf{81}}, 187-207 (1977).

\bibitem{pou2013}
A. Pouquet and R. Marino,
``Turbulence in geophysical flows and the duality of the energy flow across scales,''
Phys.\ Rev.\ Lett.\ {\bf{111}}, 234501-1-5 (2013).

\bibitem{mar2013a}
R. Marino, P. D. Mininni, D. Rosenberg, and A. Pouquet, 
``Inverse cascades in rotating stratified turbulence: Fast growth of large scales,''
Euro.\ Phys.\ Lett.\ {\bf{102}}, 44006-1-6 (2013).

\bibitem{mar2013b}
R. Marino, P. D. Mininni, D. Rosenberg, and A. Pouquet,
``Emergence of helicity in rotating stratified turbulence,''
Phys.\ Rev.\ E {\bf{87}}, 033016 (2013).

\bibitem{pel1985}
R. B. Pelz, V. Yakhot, S. A. Orszag, L. Shtilman, and E. Levich,
``Velocity--vorticity patterns in turbulent flow,''
Phys.\ Rev.\ Lett.\ {\bf{54}}, 2505-2508 (1985).

\bibitem{rog1987}
M. M. Rogers and P. Moin,
``Helicity fluctuations in incompressible turbulent flows,''
Phys.\ Fluids {\bf{30}}, 2662-2671 (1987).

\bibitem{wal1992}
J. M. Wallace, J.-L. Balint, and L. Ong,
``An experimental study of helicity density in turbulent flows,''
Phys.\ Fluids A {\bf{4}}, 2013-2026 (1992).

\bibitem{kra1974}
F. Krause and G. R\"{u}diger,
``On the Reynolds stresses in mean-field hydrodynamics I. Incompressible homogeneous isotropic turbulence,''
Astron.\ Nachr.\ {\bf{295}}, 93-99 (1974).

\bibitem{moi1983}
S. S. Moiseev, R. Z. Sagdeev, A. V. Tur, G. A. Khomenko, and V. V. Yanovsky,
``A theory of large-scale structure origination in hydrodynamic turbulence,''
Sov.\ Phys.\ JETP {\bf{53}}, 1149 (1983).

\bibitem{chk1988}
O. G. Chkhetiani and V. V. Gvaramadze,
``Generation of large-scale vortices in compressible helical turbulence,''
in {\it Proceedings of the Joint Varenna-Abastumani International School \& Workshop on Plasma Astrophysics}, held in Varenna, Italy, 24 Aug.\ - 3 Sept.\ 1988 (ESA SP-285, Vol.~I, Dec.\ 1988), pp.~363-369.

\bibitem{kho1991}
G. A. Khomenko, S. S. Moiseev, and A. V. Tur,
``The hydrodynamical alpha-effect in a compressible medium,''
J. Fluid Mech.\ {\bf{225}}, 355-369 (1991).

\bibitem{kit1994}
L. L. Kitchatinov, G. R\"{u}diger, and G. Khomenko,
``Large-scale vortices in rotating stratified disks,''
Astron.\ Astrophys.\ {\bf{287}}, 320-324 (1994).

\bibitem{fri1987}
U. Frisch, Z. S. She, and P. L. Sulem,
``Large scale flow driven by the anisotropic kinetic alpha effect,''
Physica D {\bf{28}}, 382-392 (1987).

\bibitem{gva1989}
V. V. Gvaramadze, G. A. Khomenko, and A. V. Tur,
``Large-scale vortices in helical turbulence of incompressible fluid,''
Geophys. Astrophys. Fluid Dyn.\ {\bf{46}}, 53-69 (1989).

\bibitem{chk1994}
O. G. Chkhetiany, S. S. Moiseev, and A. Petrosyan,
``The large-scale stability and self-organization in homogeneous turbulent shear flow,''
Phys. Scripta {\bf{49}}, 214-220 (1994).

\bibitem{kob1987}
T. Kobayashi and M. Yoda,
``Modified $k$--$\epsilon$ model for turbulent swirling flow in a straight pipe,''
JSME Intl.\ J. {\bf{30}}, 66-71 (1987).

\bibitem{kit1991}
O. Kitoh,
``Experimental study of turbulent swirling flow in a straight pipe,''
J. Fluid Mech.\ {\bf{225}}, 445-479 (1991).

\bibitem{ste1995}
W. Steenbergen,
``Turbulent pipe flow with swirl,''
Dissertation, Eindhoven University of Technology (1995).

\bibitem{yos1984}
A. Yoshizawa,
``Statistical analysis of the deviation of the Reynolds stress from its eddy-viscosity representation,''
Phys.\ Fluids {\bf{27}}, 1377-1387 (1984).

\bibitem{nay1973}
A. H. Nayfeh,
{\it Perturbation Methods}
(John Wiley \& Sons, Inc., New York, 1973).

\bibitem{bat1953}
G. K. Batchelor,
{\it The Theory of Homogeneous Turbulence}
(Cambridge University Press, Cambridge, 1953).

\bibitem{mon1975}
A. S. Monin and A. M. Yaglom,
{\it Statistical Fluid Mechanics: Mechanics of Turbulence}
(The MIT Press, Massachusetts, 1975).

\bibitem{lau1972}
B. E. Launder and D. B. Spalding,
{\it Lectures in Mathematical Models of Turbulence} 
(Academic Press, London, 1972).

\bibitem{rue1980}
G. R\"{u}diger,
``Reynolds stresses and differential rotation I.  On recent calculations of zonal fluxes in slowly rotating stars,''
Geophys.\ Astrophys.\ Fluid Mech.\ {\bf{16}}, 239-261 (1980).

\bibitem{rue1989}
G. R\"{u}diger,
{\it Differential Rotation and Stellar Convection: Sun and Solar-type Stars},
(Gordon and Breach Scientific Publishers, New York, 1989).

\bibitem{yok2016a}
N. Yokoi,
``Modeling helicity dissipation-rate equation,''
in {\it Progress in Turbulence VI, Springer Proceedings in Physics 165}, edited by J. Peinke {\textit{et al}}.\ DOI 10.1007/978-3-319-29130-7-17
(Springer Verlag, Heidelberg, 2016), pp.~93-96.

\bibitem{BvR01}
A. Brandenburg and B. von Rekowski,
``Astrophysical significance of the anisotropic kinetic alpha effect,''
Astron.\ Astrophys.\ {\bf{379}}, 1153-1160 (2001).

\bibitem{KR80}
F. Krause and K.-H. R\"adler,
{\it Mean-field Magnetohydrodynamics and Dynamo Theory},
(Pergamon Press, Oxford 1980).

\bibitem{yok2016b}
N. Yokoi, D. Schmitt, V. Pipin, and F. Hamba,
``A new simple dynamo model for stellar activity cycle,''
submitted to Astrophys.\ J. (2016) arXiv:1601.06348.

\bibitem{yos1990}
A. Yoshizawa,
``Self-consistent turbulent dynamo modeling of reversed field
pinches and planetary magnetic fields,''
Phys.\ Fluids B {\bf{2}}, 1589-1600 (1990).


\end{thebibliography}
\end{document}